\documentclass[aps,preprint,superscriptaddress]{revtex4-1}
\usepackage{graphicx,amsmath,amsfonts}
\begin{document}
\title{Superexchange theory of electronic polarization driven by relativistic spin-orbit interaction
at the half-filling}
\author{I. V. Solovyev}
\email{SOLOVYEV.Igor@nims.go.jp}
\affiliation{International Center for Materials Nanoarchitectonics,
National Institute for Materials Science, 1-1 Namiki, Tsukuba,
Ibaraki 305-0044, Japan}
\affiliation{Department of Theoretical Physics and Applied Mathematics, Ural Federal University,
Mira str. 19, 620002 Ekaterinburg, Russia}
\date{\today}

%Collaboration name if desired (requires use of superscriptaddress
%option in \documentclass). \noaffiliation is required (may also be
%used with the \author command).
%\collaboration can be followed by \email, \homepage, \thanks as well.
%\collaboration{}
%\noaffiliation
\date{\today}
\begin{abstract}
By applying Berry-phase theory for the effective half-filled Hubbard model,
we derive an analytical expression for the electronic polarization
driven by the relativistic spin-orbit (SO) coupling. The model itself is constructed in the Wannier basis,
using the input from the first-principles electronic structure calculations in the local-density approximation,
and then treated in the spirit of the superexchange theory.
The obtained polarization has the following form:
${\bf P}_{ij} = \boldsymbol{\epsilon}_{ji} \boldsymbol{\cal P}_{ij} \cdot [\boldsymbol{e}_i \times \boldsymbol{e}_j]$,
where $\boldsymbol{\epsilon}_{ji}$ is the direction of the bond $\langle ij \rangle$,
$\boldsymbol{e}_i$ and $\boldsymbol{e}_j$ are the directions of spins in this bond, and
$\boldsymbol{\cal P}_{ij}$ is the pseudovector containing all the information about the crystallographic symmetry
of the considered system. The expression describes the ferroelectric activity in various
magnets with noncollinear but otherwise nonpolar magnetic structures,
which would yield no polarization without SO interaction,
including the magnetoelectric (ME) effect,
caused by the ferromagnetic canting of spins in the external magnetic field, and
spin-spiral multiferroics. The abilities of this theory are demonstrated
for the the analysis of
linear ME effect in Cr$_2$O$_3$ and BiFeO$_3$ and properties
multiferroic MnWO$_4$ and $\beta$-MnO$_2$. In all considered examples, the theory perfectly describes the symmetry
properties of the induced polarization. However, in some cases,
the values of this polarization are underestimated, suggesting that other effects, besides the spin and electronic ones,
can also play an important role.
\end{abstract}

% insert suggested PACS numbers in braces on next line
% insert suggested keywords - APS authors don't need to do this
%\keywords{}
%\maketitle must follow title, authors, abstract, \pacs, and \keywords
\maketitle
\section{\label{sec:Intro} Introduction}
The relativistic spin-orbit (SO) interactions is responsible for many spectacular phenomena
in condensed matter physics,
which are widely employed in many technological applications.
Particularly,
being a natural mechanism connecting spin and orbital degrees of freedom, it provides a unique
possibility for the mutual control of various spin and lattice-related properties.
Every year, growing interest in this problem leads to the discovery new and more sophisticated schemes of
such control~\cite{SO_review}.

  One of the interesting topics is the effects of the SO coupling in noncentrosymmetric substances.
In magnetic systems, it leads to the famous antisymmetric Dzyaloshinskii-Moriya (DM) interaction
$\boldsymbol{d}_{ij} \cdot [\boldsymbol{e}_i \times \boldsymbol{e}_j]$ between
spins
in the noncentrosymmetric bond $\langle ij \rangle$, where
$\boldsymbol{e}_i$ and $\boldsymbol{e}_j$ are the directions
of these spins~\cite{Dzyaloshinskii_weakF,Moriya_weakF}.
The DM interaction is generally responsible for the noncollinear spin order. Alternatively,
in some magnetic architectures, the noncollinear alignment of spins can break the inversion symmetry,
which will be immediately manifested
in the ferroelectric (FE) activity. The classical example
of such activity is the magnetoelectric (ME) effect,
where the noncollinearity is induced by the external magnetic field~\cite{Dzyaloshinskii_ME}.
The interest in this problem has reemerged a decade ago, after the discovery of new generation of
multiferroic materials, where the inversion symmetry is broken by some complex and, in many cases,
noncollinear magnetic order~\cite{MF_review}. Nevertheless, the microscopic understanding of
mechanisms resulting in finite electric polarization in this case is still far from being complete,
even despite of significant progress in this direction.

  Historically, the first phenomenological expression
for the electric polarization, which can be induced by a noncollinear spin order, was introduced by Moriya in 1968
on the basis of general symmetry considerations~\cite{Moriya_pol}. In each magnetic bond,
such polarization has the form:
\begin{equation}
P_{ij}^a = \sum_b \mathfrak{d}_{ij}^{ab} [\boldsymbol{e}_i \times \boldsymbol{e}_j]^b,
\label{eqn:Moriya}
\end{equation}
which
is similar to the expression for DM exchange interaction, where the vector $\boldsymbol{d}_{ij}$
is replaced by the tensor $\mathfrak{d}_{ij}^{ab}$ with $a$ and $b$ denoting $x$, $y$, or $z$.

  The microscopic derivation of expression for the electric polarization,
which is driven by the relativistic SO coupling
in noncollinear magnetic substances,
was given in Ref.~\cite{KNB}. However, it should be understood that
the microscopic model considered Ref.~\cite{KNB} deals with
very special example of electronic structure of the transition-metal (TM) oxides,
consisting of the $t_{2g}$ levels with some particular scheme of filling, which are split by the SO coupling
and interact via intermediate oxygen (O) states in the single undistorted TM-O-TM bond.
Thus, the analysis is hardly to be complete.
Nevertheless, on the basis of these considerations,
the authors of Ref.~\cite{KNB} have concluded that the electric polarization should behaves as
\begin{equation}
{\bf P}_{ij} \propto \boldsymbol{\epsilon}_{ji} \times [\boldsymbol{e}_i \times \boldsymbol{e}_j],
\label{eqn:phenomenology}
\end{equation}
where $\boldsymbol{\epsilon}_{ji}$ is the unit vector in the direction of TM site $j$ relative to the TM site $i$.
It is referred to as the spin-current mechanism of the electric polarization,
which is widely used today for the analysis of experimental data~\cite{MF_review}.
Similar conclusion was drawn in Ref.~\cite{Mostovoy}, being based on the phenomenological Ginzburg-Landau theory.
The expression (\ref{eqn:phenomenology}) does not depend on the specific
crystallographic symmetry of the considered system, so that one can
have a wrong impression that the electric polarization in all noncollinear magnets should behave in a
similar way. Nevertheless, this expression is formally consistent with
the general definition~(\ref{eqn:Moriya}), given by Moriya,
and can be reduced to it by introducing the tensor
$\mathfrak{d}_{ij}^{ab} = - \varepsilon_{abc} \epsilon_{ji}^c$, where $\varepsilon_{abc}$
is the antisymmetric symbol of Levi-Civita.
It is often claimed that the microscopic mechanism responsible for such behavior
is similar to the inverse DM mechanism, proposed in Ref.~\cite{SergienkoPRB}:
similar to what how the off-centrosymmetric oxygen displacement in the bond TM-O-TM gives rise
to the noncollinear alignment of spins~\cite{Dzyaloshinskii_weakF,Moriya_weakF}, one can expect the opposite
(magnetostrictive-like) effect,
where the noncollinear magnetic alignment should lead to the off-centrosymmetric atomic displacement
responsible for the additional magnetic energy gain:
$\sum_{ij} \boldsymbol{d}_{ij} \cdot [\boldsymbol{e}_i \times \boldsymbol{e}_j]$.
However, it should be understood
that these mechanisms are quite different (though complementary to each other):
Ref.~\cite{KNB} deals with the
purely electronic effect, while Ref.~\cite{SergienkoPRB} deals with the lattice effect.

  The most rigorous theoretical basis for the analysis of electronic polarization
is provided by the Berry-phase theory, which relates the polarization with the expectation value of the
position operator in the state specified by localized Wannier functions~\cite{KSV,Resta,W_review}:
\begin{equation}
{\bf P} = - \frac{e}{V}
\int {\bf r} \, w^2({\bf r}) \, d {\bf r},
\label{eqn:PWannier}
\end{equation}
where $-$$e < 0$ is the electron charge,
$V$ is the unit-cell volume, and $w^2({\bf r}) = \sum_{n=1}^M |W_n({\bf r})|^2$ is the total weight of Wannier functions
for the $M$ occupied states. Each Wannier function is centered near certain site of the lattice and can have tails
spreading to the neighboring sites. The relative weight of these tails depends on the magnetic state.
This is how the Wannier function
bears the information about the magnetic configuration at the
neighboring sites.
Thus, the understanding of magnetic-state dependence of the electronic polarization
is essentially the understanding of how the magnetic order and relativistic SO interaction leads to
the asymmetric deformation of the Wannier functions around each magnetic site~\cite{Barone,DEpol,Yamauchi_rew}.
It should not be confused with
the asymmetric distribution of the electron density at each magnetic site, because
the electron density is a superposition of the weights of the Wannier functions
centered at this and neighboring sites, which can lead to the incorrect answer~\cite{KSV,Resta}.

  In our previous work~\cite{DEpol} we have applied this strategy to the analysis of electronic polarization
caused by the nonrelativistic double exchange mechanism. In that case, competing magnetic interactions
of both relativistic and nonrelativistic origin result in highly asymmetric magnetic structure,
which breaks the inversion symmetry. The SO interaction plays an important role in
this asymmetry: for instance, it is responsible for the single-ion anisotropy, which deforms the
homogeneous spin-spiral texture in multiferroic manganites~\cite{DEpol,PRB11}
(the so-called bunching effect~\cite{REbunching}).
This deformation gives rise to the polarization
${\bf P}_{ij} \propto (\boldsymbol{e}_i \cdot \boldsymbol{e}_j)$, which depends on the SO coupling
only indirectly, via the noncentrosymmetric distribution of the directions of spins,
while the proportionality coefficient between ${\bf P}_{ij}$ and $(\boldsymbol{e}_i \cdot \boldsymbol{e}_j)$
does not depend on the SO coupling.
This double exchange mechanism has allowed us to rationalize many aspects of the behavior of electric
polarization in multiferroic manganites~\cite{DEpol}.

  In this article we consider the proper spin-current mechanism. In some sense, the situations is the
opposite to the double exchange mechanism, considered in Ref.~\cite{DEpol}. Namely, we will deal with some
noncollinear magnetic structures, which are
stabilized by nonrelativistic means: it can be either the spin-spiral structure arising from the
competition of several isotropic exchange interactions or a canted spin structure, inherent to the
ME effect, where the
the collinear antiferromagnetic (AFM) order is deformed by the external magnetic field. Without SO coupling
all these magnetic structures can be transformed to themselves by combining the spacial inversion with some
appropriate rotation of the spin system as the whole~\cite{PRB13}. Therefore, the electric polarization
will vanish. Nevertheless, the situation may change after switching on the SO coupling, which does not deform
the spin texture itself (or, at least, such deformation can be neglected), but can deform the Wannier functions,
resulting in their asymmetry and finite electronic polarization.

  Our analysis will be applied to the effective Hubbard model derived
from the first-principles electronic structure calculations and using the local-density approximation (LDA)
as the starting point for such derivation~\cite{review2008}.
We consider the simplest case of the half-filling, which also allows us to get rid of additional
complications related to the orbital degrees of freedom. Furthermore,
the on-site Coulomb repulsion is the largest parameter in our model, so that other parameters
can be treated as a perturbation in the spirit of the superexchange (SE) theory~\cite{PWA}.
We will use this strategy in order to derive an analytical expression for the DM exchange interactions and
electronic polarization. We will show that the correct expression for electronic
polarization, which driven by the spin-current mechanism
in the framework of the Berry-phase theory~\cite{KSV,Resta}, has the following form:
\begin{equation}
{\bf P}_{ij} = \boldsymbol{\epsilon}_{ji} \boldsymbol{\cal P}_{ij} \cdot [\boldsymbol{e}_i \times \boldsymbol{e}_j],
\label{eqn:Ppartial2}
\end{equation}
where the pseudovector $\boldsymbol{\cal P}_{ij}$ contains all the information about the individual symmetry
of the lattice.
Thus, there is at least one important addition to the phenomenological expression (\ref{eqn:phenomenology}):
the polarization does depend on the symmetry of the lattice, as it should be.
Moreover, the functional dependence is different
and there is no direct coupling between $\boldsymbol{\epsilon}_{ji}$
and $[\boldsymbol{e}_i \times \boldsymbol{e}_j]$.
Furthermore, by defining $\mathfrak{d}_{ij}^{ab} = \epsilon_{ji}^a {\cal P}_{ij}^b$,
it is also straightforward to see the form of this expression is consistent
with Eq.~(\ref{eqn:Moriya}), proposed by Moriya~\cite{Moriya_pol}.
We will show that this expression is very general and
describes not only the behavior of polarization in the spin-spiral magnets, but
also the ME effect,
caused by the ferromagnetic (FM) canting of spins in
otherwise collinear AFM states of a special symmetry, which is not captured
by phenomenological Eq.~(\ref{eqn:phenomenology}).

  Another important issue is whether the spin-current mechanism alone is able to reproduce experimental values of
the ME effect and electric polarization in real materials. Additional mechanisms, which are widely
discussed in the literature, are the lattice deformation~\cite{Malashevich2008,Bousquet,Malashevich}
(in line with the proposal~\cite{SergienkoPRB}),
orbital contribution to the ME coupling~\cite{Malashevich,Scaramucci}, and hidden deformation
of the magnetic texture with broken
inversion symmetry~\cite{PRB11,PRB13}. By using realistic model, derived from the
first-principles calculations, we will show that the situation can be very different: In some cases,
the spin-current mechanism alone reproduces the experimental polarization reasonably well. In other cases
(e.g., in Cr$_2$O$_3$), it captures only the symmetry properties of the polarization, while the numerical values
can be off by several order of magnitude, suggesting the importance of
other mechanisms~\cite{Bousquet,Malashevich,Scaramucci}.

  The rest of the article is organized as follows. In Sec.~\ref{sec:Method} we will present our formalism
based on the SE theory, which is applied to antisymmetric DM exchange interactions and electric polarization in
Secs.~\ref{sec:exchange} and \ref{sec:polarization}, respectively.
The details of these derivations are given in the Supplemental Material~\cite{SM}.
In Sec.~\ref{sec:Results}
we will consider practical applications of this formalism to the linear ME effect in Cr$_2$O$_3$ and
BiFeO$_3$ (Secs.~\ref{sec:Cr2O3} and \ref{sec:BiFeO3}, respectively) and FE
activity caused by the spin-spiral order in multiferroic MnWO$_4$ and $\beta$-MnO$_2$
(Sec.~\ref{sec:MnWO4} and \ref{sec:MnO2}, respectively).
Finally, in Sec.~\ref{sec:conc}, we will summarize our work.

\section{\label{sec:Method} Formalism}
In this section we will sketch the main details of derivation of analytical expressions
for the DM exchange interactions and electric polarization, following the SE theory
in the lowest order of perturbation with respect to the transfer integrals
$\hat{t}_{ij}$~\cite{PWA}. The technical details can be found in the Supplemental Material~\cite{SM}.
The simplest microscopic model, capturing the physics of the spin-current mechanism, reads $\hat{H} = \hat{h} + \hat{t}$,
where $\hat{h} \equiv \hat{h}_{\rm ex} + \hat{h}_{\rm cf} + \hat{h}_{\rm so}$ is the on-site part,
including the interaction $\hat{h}_{\rm ex} = \frac{U}{2} \boldsymbol{e} \cdot \hat{\boldsymbol{\sigma}}$
with the internal exchange field
in the direction $\boldsymbol{e} = (\sin \theta \cos \phi, \sin \theta \sin \phi, \cos \theta)$
($\hat{\boldsymbol{\sigma}}$ being the vector of Pauli matrices),
the crystal-field splitting $\hat{h}_{\rm cf}$,
and the SO interaction $\hat{h}_{\rm so} = \frac{\xi}{2} \hat{\bf L} \cdot \hat{\boldsymbol{\sigma}}$,
while $\hat{t} \equiv [ \hat{t}_{ij} ]$ is the inter-site part. More specifically, $\hat{H}$ can be viewed as a
mean-field Hamiltonian (for instance,
the one obtained from the solution of the Hubbard model in the Hartree-Fock approximation),
where $\hat{h}_{\rm ex}$ describes the averaged exchange splitting for the half-filled ionic shell, driven by the
effective interaction $U$,
and the crystal field $\hat{h}_{\rm cf}$ also includes
the effects of nonsphericity of the Coulomb and exchange potential.
The form of $\hat{h}_{\rm ex}$ implies that spins are decoupled from the orbital degrees of freedom, which do not adjust
the reorientation of spins. Thus, what we consider here is the canonical ``spin-current'' model, which
include direct contributions of the orbital magnetization in neither
DM interactions nor electric polarization.
Parameters of such microscopic model, formulated in the Wannier basis~\cite{W_review}, can be derived
from the first-principles electronic structure calculations~\cite{review2008}.
For practical purposes we use
the linear muffin-tin orbital (LMTO) method~\cite{LMTO}.

  The basic idea of the SE theory is to start from the atomic limit and treat $\hat{h}_{\rm so}$ and $\hat{t}$
as a perturbation. Then, the wavefunctions of $\hat{H}_0 = \hat{h}_{\rm ex} + \hat{h}_{\rm cf}$ for the
occupied the occupied ($-$) and unoccupied ($+$) spin states are given by
$$
| \Psi^- \rangle =
\left(
\begin{array}{l}
- \sin \frac{\theta}{2} e^{-i \phi} \\
\phantom{-} \cos \frac{\theta}{2} \\
\end{array}
\right) | \Psi \rangle
$$
and
$$
| \Psi^+ \rangle =
\left(
\begin{array}{l}
\cos \frac{\theta}{2} \\
\sin \frac{\theta}{2} e^{ i \phi} \\
\end{array}
\right) | \Psi \rangle,
$$
respectively, where $| \Psi \rangle$ is the column of eigenvectors of $\hat{h}_{\rm cf}$
with the eigenvalues $\{ \varepsilon_n \}$. More specifically, $| \Psi \rangle$ is the
$M$-dimensional vector in the subspace of orbital states, while $| \Psi^{\pm} \rangle$ are $2M$-dimensional vectors in the space of
spin and orbital states. Then, corresponding eigenvectors in the first order of the SO interaction
will be given by
$$
| \tilde{\Psi}^- \rangle = | \Psi^- \rangle - \bar{\xi} | \Psi^+ \rangle
\langle \Psi^+ | \left( \hat{\bf L} - [ \hat{\bar{h}}_{\rm cf}, \hat{\bf L}  ]
\right) \cdot \hat{\bf S} | \Psi^- \rangle
$$
and
$$
| \tilde{\Psi}^+ \rangle = | \Psi^+ \rangle + \bar{\xi} | \Psi^- \rangle
\langle \Psi^- | \left( \hat{\bf L} + [ \hat{\bar{h}}_{\rm cf}, \hat{\bf L}  ]
\right) \cdot \hat{\bf S} | \Psi^+ \rangle,
$$
where
$\bar{\xi}=\xi/U$, $\hat{\bar{h}}_{\rm cf} = \hat{h}_{\rm cf}/U$, and $[\hat{A},\hat{B}] = \hat{A} \hat{B} - \hat{B} \hat{A}$.
Moreover, in the conventional perturbation theory expression,
we further expand $(\varepsilon_n - \varepsilon_m \pm U)^{-1}$
with respect to $\hat{\bar{h}}_{\rm cf}$. Then,
the first terms in $(\dots)$ correspond to $\hat{h}_{\rm cf} = const$ (the constant energy shift), while the second
terms appear in the first order of $\hat{\bar{h}}_{\rm cf}$. In practical calculations, we use the effective $\xi$,
which also incorporates the change of the Coulomb and exchange potential in the first order of the SO interaction,
as obtained in the self-consistent linear response (SCLR) theory~\cite{SCLR}.

\subsection{\label{sec:exchange} Exchange Interactions}
The exchange interactions in the bond $\langle ij \rangle$ describe the energy change $\delta E_{ij}$
in the second order of $\hat{\bar{t}}_{ij} = \hat{t}_{ij}/U$,
where the transfer integrals connect the occupied and unoccupied states
of the sites $i$ and $j$:
$$
\delta E_{ij} \simeq -U  \langle \tilde{\Psi}^-_i | \hat{\bar{t}}_{ij} +
\frac{1}{2} [\hat{\bar{h}}_{\rm cf}, \hat{\bar{t}}_{ij} ]  | \tilde{\Psi}^+_j \rangle
\langle \tilde{\Psi}^+_j | \hat{\bar{t}}_{ji} - \frac{1}{2} [\hat{\bar{h}}_{\rm cf}, \hat{\bar{t}}_{ji} ] | \tilde{\Psi}^-_i \rangle
+ (i \leftrightarrow j).
$$
This expression is also valid in the first order of $\hat{\bar{h}}_{\rm cf}$.
Then, after tedious but rather straightforward algebra,
it can be rearranged as (see Ref.~\cite{SM} for details)
\begin{equation}
\delta E_{ij} \simeq J_{ij} \left( 1 - \boldsymbol{e}_i \cdot \boldsymbol{e}_j \right) + \boldsymbol{d}_{ij}
\cdot [\boldsymbol{e}_i \times \boldsymbol{e}_j],
\label{eqn:Eij}
\end{equation}
where
\begin{equation}
J_{ij} = - U {\rm Tr}_L \left\{ \hat{\bar{t}}_{ij} \hat{\bar{t}}_{ji} \right\}
\label{eqn:Jij}
\end{equation}
is the isotropic exchange coupling, which does not depend on the SO intercation, and
\begin{equation}
\boldsymbol{d}_{ij} = \xi {\rm Tr}_L \left\{
\hat{\bar{t}}_{ij} [ [\hat{\bar{h}}_{\rm cf},i \hat{\bf L}], \hat{\bar{t}}_{ji} ]
\right\}
\label{eqn:dij}
\end{equation}
is the DM coupling, which appears in the first order of the SO interaction.
Other exchange interactions, including the
symmetric anisotropic one,
appear only in higher orders of the SO coupling.
${\rm Tr}_L$ in Eqs.~(\ref{eqn:Jij}) and (\ref{eqn:dij}) denotes the trace over $M$ orbital indices.

  Finally, we note the following properties: \\
(i) $\boldsymbol{d}_{ij}$ is the antisymmetric pseudovector:
$\hat{I} \boldsymbol{d}_{ij} = \boldsymbol{d}_{ij}$, and $\boldsymbol{d}_{ji} = -\boldsymbol{d}_{ij}$; \\
(ii) The values of the DM interactions depend on the crystal-field splitting
and vanish when $\hat{\bar{h}}_{\rm cf} = const$. Then, Eq.~(\ref{eqn:dij}) can be
interpreted in the following way: since $[\hat{\bar{h}}_{\rm cf},i \hat{\bf L}]$ is the measure of
unquenched orbital magnetization (or the observable orbital magnetization in the presence of the crystal field), the
DM interactions $\boldsymbol{d}_{ij}$ is a probe of the orbital magnetization at the site $j$ by the electron hoppings
from the site $i$ (and vice versa).

\subsection{\label{sec:polarization} Electronic Polarization}
We start with the general expression for the electric polarization (\ref{eqn:PWannier})
in terms of the Wannier functions for the occupied states.
Moreover, we adopt it for the lattice model and assume that
all weights of $w$ are localized in the lattice points: i.e., if $w_i$ are occupied Wannier functions
centered at the site $i$,
their weights are distributed as
$$
w_i^2({\bf r}) = \sum_j w_{ij}^2 \, \delta({\bf r} - \Delta \boldsymbol{\tau}_{ji}),
$$
where $\Delta \boldsymbol{\tau}_{ji} = \boldsymbol{R}_j - \boldsymbol{R}_i$ is the position of the site $j$ relative to the site $i$.
Then, the electronic polarization (\ref{eqn:PWannier})
can be related to the asymmetric transfer of the weights of the Wannier functions in each bond~\cite{DEpol}:
\begin{equation}
{\bf P} = \frac{1}{2} \sum_{ij} {\bf P}_{ij},
\label{eqn:Psum}
\end{equation}
where
\begin{equation}
{\bf P}_{ij} = -\frac{e \Delta \boldsymbol{\tau}_{ji}}{V} \left( w_{ij} - w_{ji} \right).
\label{eqn:Pdef}
\end{equation}
In the SE theory, the quantities $w_{ij}$ are evaluated in the first order of perturbation theory for the Wannier functions
with respect to $\hat{\bar{t}}_{ij}$, starting from the atomic limit:
$$
w_{ij}  \simeq
| \langle \tilde{\Psi}^+_j | \hat{\bar{t}}_{ji} - [\hat{\bar{h}}_{\rm cf}, \hat{\bar{t}}_{ji} ]  | \tilde{\Psi}^-_i \rangle |^2  .
$$
Then, using tedious but rather straightforward algebra,
one can obtain the following expression for ${\bf P}_{ij}$ (see Ref.~\cite{SM} for details):
\begin{equation}
{\bf P}_{ij} = \frac{e \Delta \boldsymbol{\tau}_{ji}}{V} \bar{\xi} {\rm Tr}_L \left\{
[\hat{\bar{h}}_{\rm cf},\hat{\bar{t}}_{ij}] [ [\hat{\bar{h}}_{\rm cf},i \hat{\bf L}], \hat{\bar{t}}_{ji} ] +
[\hat{\bar{h}}_{\rm cf},\hat{\bar{t}}_{ji}] [ [\hat{\bar{h}}_{\rm cf},i \hat{\bf L}], \hat{\bar{t}}_{ij} ]
\right\} \cdot [\boldsymbol{e}_i \times \boldsymbol{e}_j],
\label{eqn:Ppartial}
\end{equation}
which can be further rearranged as Eq.~(\ref{eqn:Ppartial2})
whith $\boldsymbol{\epsilon}_{ji} = \frac{\Delta \boldsymbol{\tau}_{ji} }{ | \Delta \boldsymbol{\tau}_{ji} | }$ and
$$
\boldsymbol{\cal P}_{ij} = \frac{e | \Delta \boldsymbol{\tau}_{ji}|}{V} \bar{\xi} {\rm Tr}_L \left\{
[\hat{\bar{h}}_{\rm cf},\hat{\bar{t}}_{ij}] [ [\hat{\bar{h}}_{\rm cf},i \hat{\bf L}], \hat{\bar{t}}_{ji} ] +
[\hat{\bar{h}}_{\rm cf},\hat{\bar{t}}_{ji}] [ [\hat{\bar{h}}_{\rm cf},i \hat{\bf L}], \hat{\bar{t}}_{ij} ]
\right\}.
$$

  Thus, we note the following: \\
(i) Unlike $\boldsymbol{d}_{ij}$, $\boldsymbol{\cal P}_{ij}$ is the \textit{symmetric} pseudovector:
$\hat{I} \boldsymbol{\cal P}_{ij} = \boldsymbol{\cal P}_{ij}$, \textit{while}
$\boldsymbol{\cal P}_{ji} = \boldsymbol{\cal P}_{ij}$, where the latter property
comes from the definition of ${\bf P}_{ij}$ [Eq.~(\ref{eqn:Pdef})];\\
(ii) Similar to the DM interactions,
the electronic polarization crucially depends on $\hat{\bar{h}}_{\rm cf}$
and vanishes when $\hat{\bar{h}}_{\rm cf} = const$;\\
(iii) There is a fundamental difference from phenomenological expression (\ref{eqn:phenomenology})~\cite{KNB,Mostovoy}.
Namely, the spin-dependent cross product $[\boldsymbol{e}_i \times \boldsymbol{e}_j]$ does not couple
directly to $\boldsymbol{\epsilon}_{ji}$. It couples to the pseudovector $\boldsymbol{\cal P}_{ji}$,
which contains all the information about particular
crystallographic symmetry of the considered system. The directional dependence of
${\bf P}$ is specified by the vectors $\boldsymbol{\epsilon}_{ji}$, which are modulated by the scalar products
$\boldsymbol{\cal P}_{ij} \cdot [\boldsymbol{e}_i \times \boldsymbol{e}_j]$. This important
addition will allow us to resolve several controversies related to the symmetry properties
of the electric polarization induced by the noncollinear magnetic alignment.

\section{\label{sec:Results} Results and Discussions}

\subsection{\label{sec:Cr2O3} Perpendicular Magnetoelectric Effect in Cr$_2$O$_3$}
\begin{figure}[tbp]
\begin{center}
\includegraphics[width=10cm]{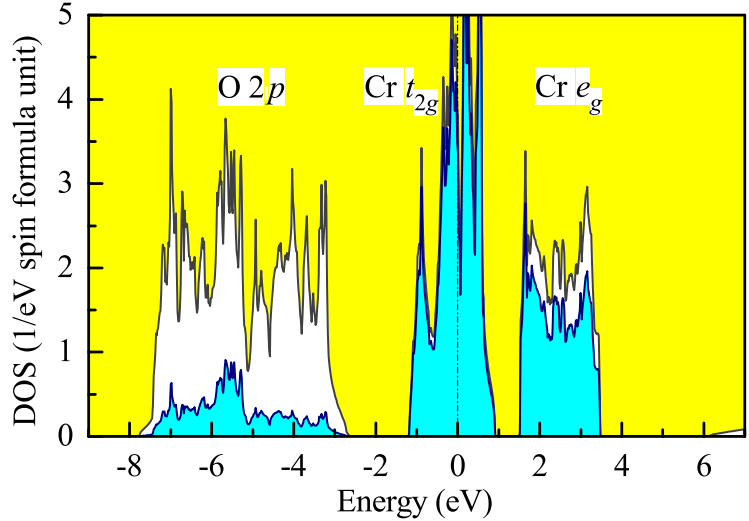}
\end{center}
\caption{(Color online)
Total and partial densities of states of Cr$_2$O$_3$ in the local density approximation.
The shaded light (blue) area shows contributions of the Cr$3d$ states.
Positions of the main bands are indicated by symbols. The Fermi level is at zero energy (shown by dot-dashed line).}
\label{fig.Cr2O3DOS}
\end{figure}

We start our discussion with the canonical example of ME effect in antiferromagnetic Cr$_2$O$_3$~\cite{Dzyaloshinskii_ME},
which crystallizes in the corundum structure (the space group is $R\bar{3}c$)~\cite{Cr2O3_structure}.
The formal configuration of the Cr-ions in Cr$_2$O$_3$ is $3d^3$.
According to the electronic structure calculations in LDA, Cr$_2$O$_3$
has well isolated Cr $t_{2g}$ band near the Fermi level, which accommodates 3 electrons per one Cr site (Fig.~\ref{fig.Cr2O3DOS}).
Thus, as a first approximation,
we consider the simplest $t_{2g}$ model at the half filling
and try to apply this model for the analysis of the ME effect in Cr$_2$O$_3$.
The model itself has been constructed in the basis of
Wannier functions, and the parameters of this model have been derived as
described in Ref.~\cite{review2008}. The obtained transfer integrals and the crystal-field splitting
perfectly reproduce Cr $t_{2g}$ band structure in LDA. The matrix of
screened Coulomb interactions, evaluated in the framework of constrained random-phase approximation (RPA)~\cite{Ferdi04},
can be approximated in terms of two
Kanamori parameters~\cite{Kanamori}: the intraorbital Coulomb repulsion ${\cal U} = 3.15$ eV and the
exchange interaction ${\cal J} = 0.67$ eV. Then, the effective interaction responsible for the
intraatomic exchange splitting between the minority- and majority-spin states can be evaluated
$U = {\cal U} + 2{\cal J}$.
The crystal-field splitting of atomic $t_{2g}$ levels is about $100$ meV~\cite{PRB06}.
Other parameters can be found elsewhere~\cite{footnote2}. As we will see below, the model
has serious limitations for the quantitative description of the ME effect in Cr$_2$O$_3$.
Nevertheless, we consider it for the explanatory purposes.

\begin{figure}[tbp]
\begin{center}
\includegraphics[width=15cm]{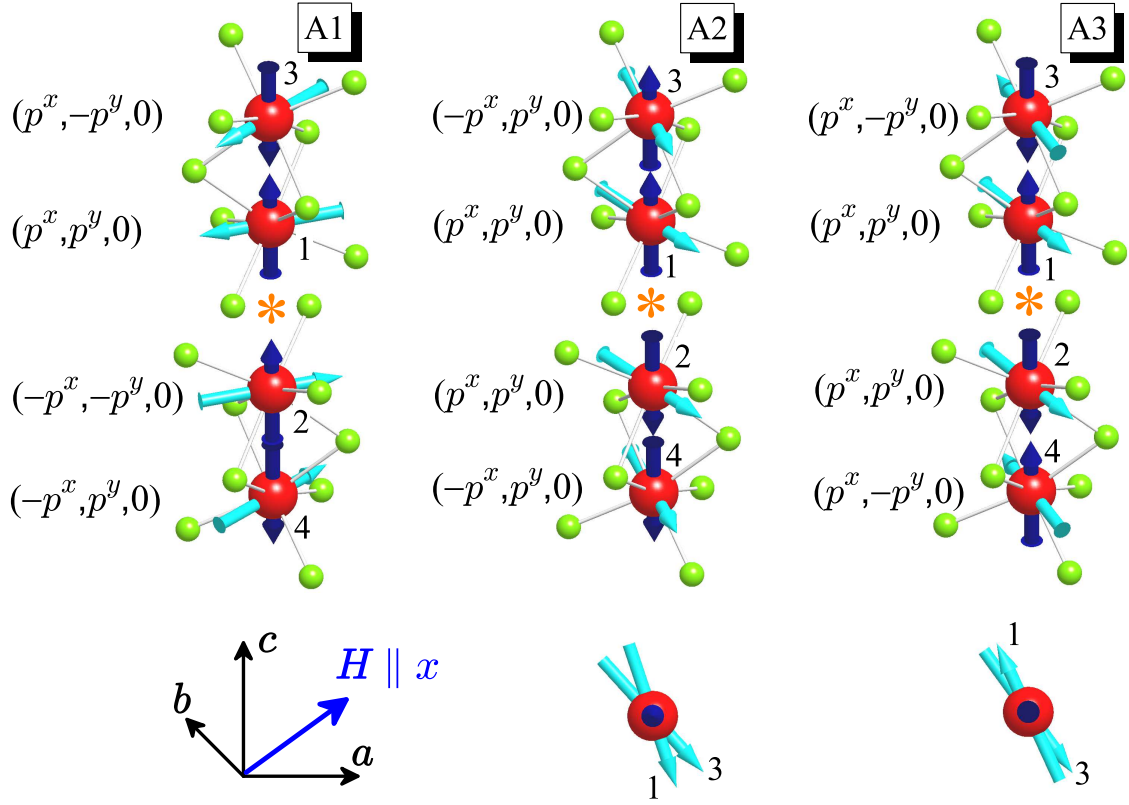}
\end{center}
\caption{(Color online)
Directions of electronic polarization at four Cr sites in the primitive cell of Cr$_2$O$_3$, which is induced
by the ferromagnetic canting of spins along the
$\boldsymbol{x}$ axis
in three possible antiferromagnetic structures. The directions of spins are denoted by the blue (dark) arrows.
The directions of electronic polarization are denoted by the cyan (light) arrows.
The Cr atoms are indicated by the big red spheres and the neighboring oxygen atoms are indicated by the small green spheres.
The inversion center is indicated by $*$. The upper panel is the side view, while the lower panel is the top view.
$\boldsymbol{a}$, $\boldsymbol{b}$, and $\boldsymbol{c}$ denote the directions of
hexagonal lattice vectors, and $\boldsymbol{H}$
denotes the external magnetic field inducing the ferromagnetic canting of spins along $\boldsymbol{x}$.
The notations $(\pm p^x, \pm p^y, 0)$ explain the symmetry properties of the induced polarization vectors $\partial {\bf P}/ \partial e^x$ at four Cr sites. The numerical values of
$(p^x,p^y)$ are  $(-0.08, 0.02)$, $(0.12,-0.57)$, and $(0.08,-0.79)$
$\mu {\rm C}/{\rm m}^2$ for the antiferromagnetic structures A1, A2, and A3, respectively.}
\label{fig.Cr2O3chain}
\end{figure}

  The corundum structure of Cr$_2$O$_3$
has four interconnected Cr sublattices,
which can be antiferromagnetically arranged as A1, A2, and A3 (see Fig.~\ref{fig.Cr2O3chain}).
Among them, the magnetic space group of A1 contains the spacial inversion $\hat{I}$ as it is, while in A2 and A3 $\hat{I}$ is combined with the
time reversal $\hat{T}$. Thus, the A1 structure allows for the weak ferromagnetism~\cite{Dzyaloshinskii_weakF},
while A2 and A3 are expected to exhibit the perpendicular ME effect, when the AFM structure is deformed by the
external magnetic field~\cite{Dzyaloshinskii_ME} as explained in Fig.~\ref{fig.Cr2O3chain}.
The magnetic ground state of Cr$_2$O$_3$ is A3, which was also confirmed by our calculations. The directions of
magnetic moments are parallel to $\boldsymbol{c}=\boldsymbol{z}$.

\begin{figure}[tbp]
\begin{center}
\includegraphics[width=10cm]{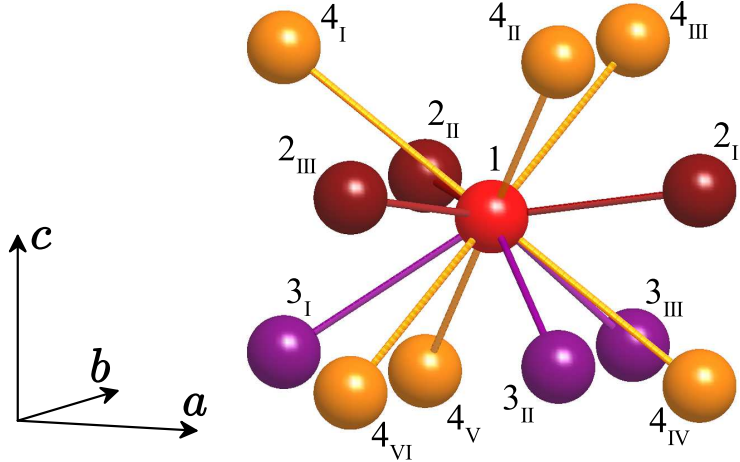}
\end{center}
\caption{(Color online)
Fragment of the crystal structure of Cr$_2$O$_3$: central Cr site of the type 1 and several coordinations spheres of
the neighboring Cr sites of the types 2, 3, and 4 with the notations of their atomic positions.
$\boldsymbol{a}$, $\boldsymbol{b}$, and $\boldsymbol{c}$ denote the directions of hexagonal lattice vectors.}
\label{fig.Cr2O3cluster}
\end{figure}

  Eqs.~(\ref{eqn:Psum}) and (\ref{eqn:Ppartial}) allow us to rationalize the behavior of
electronic polarization by separating the
contributions
of atomic pairs around each Cr site. Around site 1, the largest contributions to ${\bf P}$ comes from the atomic pairs
in three coordinations spheres, formed by the atoms 2, 3, and 4,
which are displayed in Fig.~\ref{fig.Cr2O3cluster}, and where the notations of atomic types is the same as in Fig.~\ref{fig.Cr2O3chain}.

  First, we note that the FM bond will not contribute to the
perpendicular ME effect: even in the external field $\boldsymbol{H}$ such spins
remain ferromagnetically aligned and, therefore, the cross product $[\boldsymbol{e}_i \times \boldsymbol{e}_j]$
will vanish. Moreover, in the case of perpendicular ME effect,
the cross products $[\boldsymbol{e}_i \times \boldsymbol{e}_j]$ will be the same in all equivalent bonds.

  Another important aspect is the symmetry.
In order to estimate the ME coupling constant, we first evaluate the parameters
$\boldsymbol{\cal P}_{ij}$, which obey
the symmetry properties of the $R\bar{3}c$ group and contains all the information about the individual symmetry
of the Cr$_2$O$_3$ lattice.
For instance, the pseudovectors $\boldsymbol{\cal P}_{ij}$
in the nearest-neighbor (NN) bond $\langle 13 \rangle$ (and equivalent to it bonds),
parallel to the $\boldsymbol{c}$ axis, will vanish due to the
joint effect of the threefold rotation and the glade reflection, which transform this bond to itself.
Then, for the bonds $\langle 12_{\rm I} \rangle$, $\langle 13_{\rm I} \rangle$, and $\langle 14_{\rm I} \rangle$
(see Fig.~\ref{fig.Cr2O3cluster})
the calculated parameters $\boldsymbol{\cal P}_{ij}$
are $(-0.066,0.007,0)$, $(0,0.009,0)$, and $(-0.002,0.010,0)$, respectively (in $\mu {\rm C}/{\rm m}^2$).
The parameters for other bonds can be obtained from $\boldsymbol{\cal P}_{12_{\rm I}}$, $\boldsymbol{\cal P}_{13_{\rm I}}$, and
$\boldsymbol{\cal P}_{14_{\rm I}}$ using the symmetry operations of the $R\bar{3}c$ group. Moreover, due to the
threefold rotation about $\boldsymbol{c}$,
we will have the following property: $\sum_j \boldsymbol{\cal P}_{ij} = 0$. Nevertheless,
the combination $\sum_j \boldsymbol{\epsilon}_{ji} \boldsymbol{\cal P}_{ij}$,
which specifies the value of the electronic polarization (\ref{eqn:Ppartial2}), can be finite,
which will lead to the finite ME effect. All these properties do not depend on the
type of the AFM  order and will hold for A1, A2, and A3.

  Then, we consider the behavior of polarization vectors
$\boldsymbol{p}_i = \sum_j \partial \boldsymbol{P}_{ij} / \partial e^x$,
induced by the FM canting of spins along $\boldsymbol{x}$ and
associated with each magnetic site for different types of the AFM order
(see Fig.~\ref{fig.Cr2O3chain}). In the A1 phase,
due to the FM alignment of spins in
the bonds connecting the atomic types 1 and 2 (3 and 4),
these bonds will not contribute to $\boldsymbol{p}_i$. Thus, one have to consider all possible connections of the sites 1 and 2
with the sites 3 and 4. Moreover, since the sites 1 and 2 (3 and 4) are transformed to each other by $\hat{I}$
without flipping the directions of spins,
we will have:
$\hat{I} \boldsymbol{\epsilon}_{ji} \boldsymbol{\cal P}_{ij} = - \boldsymbol{\epsilon}_{j'i'} \boldsymbol{\cal P}_{i'j'}$
but $[\boldsymbol{e}_i \times \boldsymbol{e}_j] = [\boldsymbol{e}_{i'} \times \boldsymbol{e}_{j'}]$,
and, therefore,
$\boldsymbol{p}_2 = - \boldsymbol{p}_1$ ($\boldsymbol{p}_4 = - \boldsymbol{p}_3$),
where $i'$ ($j'$) is the inversion image of $i$ ($j$).
Thus, as expected~\cite{Dzyaloshinskii_ME,Dzyaloshinskii_weakF}, the FM canting of spins in
the phase A1 will lead to the antiferroelectric behavior with no net polarization.
Our analysis provides a transparent microscopic explanation of this effect.

  In the phases A2 and A3, however, the spins 1 and 2 (3 and 4) are coupled antiferromagnetically.
Therefore, these bonds will contribute to $\boldsymbol{p}_i$. Moreover, in addition to
$\hat{I} \boldsymbol{\epsilon}_{ji} \boldsymbol{\cal P}_{ij} = - \boldsymbol{\epsilon}_{j'i'} \boldsymbol{\cal P}_{i'j'}$,
in this case we will have:
$[\boldsymbol{e}_i \times \boldsymbol{e}_j] = -[\boldsymbol{e}_{i'} \times \boldsymbol{e}_{j'}]$
(due to the $\hat{I}\hat{T}$ symmetry of the phases A2 and A3, the inversion will also flip the directions of spins)
and, therefore,
$\boldsymbol{p}_2 = \boldsymbol{p}_1$ ($\boldsymbol{p}_4 = \boldsymbol{p}_3$). This is a microscopic explanation
of the ME effect, which is expected in the phases A2 and A3.

  The direction of polarization, however, requires additional symmetry considerations, and this is the point
where the external field $\boldsymbol{H}$ comes into play.
For instance, if (without field) all $\boldsymbol{e}_i$ are parallel to $\boldsymbol{z}$ and
the field $\boldsymbol{H}$ is parallel to $\boldsymbol{x}$ (see Fig.~\ref{fig.Cr2O3chain}),
$[\boldsymbol{e}_i \times \boldsymbol{e}_j]$ will be parallel to $\boldsymbol{y}$ and,
according to Eq.~(\ref{eqn:Ppartial2}), we have to consider the behavior of ${\cal P}_{ij}^y$
and $\boldsymbol{\epsilon}_{ji}$
under the
glade reflection $\{ \hat{m}_y | \boldsymbol{c}/2 \}$
($\hat{m}_y$ being the mirror reflection $y \rightarrow -$$y$), connecting the sites 1 and 4 (2 and 3).
In the A2 phase, this transformation is combined with $\hat{T}$ and, therefore, will
additionally flip the direction of spins.
Then, it is straightforward to show
(similar to the above considerations for $\hat{I}$)
that $\{ \hat{m}_y | \boldsymbol{c}/2 \}$
leads to the the additional symmetry properties:
$p_1^x = - p_4^x$ and $p_1^y = p_4^y$ ($p_1^x = p_4^x$ and $p_1^y = - p_4^y$) for A2 (A3).
This explains why ${\bf P}$ in A2 and A3 will be parallel to, respectively, $\boldsymbol{y}$ and $\boldsymbol{x}$.

  These behavior cannot be properly
described by the phenomenological expression (\ref{eqn:phenomenology})~\cite{KNB,Mostovoy}:
in the case of ME effect, the cross product $[\boldsymbol{e}_i \times \boldsymbol{e}_j]$ is the same for all equivalent bonds.
Then, the bonds $\langle 14_{\rm I} \rangle$ - $\langle 14_{\rm VI} \rangle$ will not contribute to ${\bf P}$ because
$\sum_j \boldsymbol{\epsilon}_{ji} = 0$ (see Fig.~\ref{fig.Cr2O3cluster}). For other types of bonds
$\sum_j \boldsymbol{\epsilon}_{ji}$ is finite and parallel to $\boldsymbol{z}$.
Therefore, according to Eq.~(\ref{eqn:phenomenology}), for $\boldsymbol{H} || \boldsymbol{x}$
the induced polarization should be always parallel to $\boldsymbol{x}$.
This could explain the direction of the polarization in the A3 phase, but not in the A2 one.

  Finally, we evaluate the matrix element of the ME tensor $\alpha_\perp$ using
the numerical value of $p_x = 0.08$ $\mu {\rm C}/{\rm m}^2$ for the A3 phase and the chain rule:
$\alpha_\perp \equiv \frac{\partial P^x}{\partial H^x} = \frac{\partial P^x}{\partial e^x} \frac{\partial e^x}{\partial H^x}$,
where $\frac{\partial e^x}{\partial H^x}$ is estimated using parameters of the Heisenberg model
$E_{\rm H} = - \sum_{i > j} J_{ij} \boldsymbol{e}_i \cdot \boldsymbol{e}_j$, obtained in the theory of
infinitesimal spin rotations~\cite{review2008,JHeisenberg},
as $\frac{\partial e^x}{\partial H^x} = - \frac{M}{J_0}$
($M \approx 3$ $\mu_{\rm B}$ being the spin magnetic moment and $J_0 = \sum_j J_{ij} \approx -116$ meV) \cite{footnote1}.
It should be noted that for half-filled Mott insulators, the orbitals degrees of freedom and inactive
and parameters of exchange interactions obtained in the SE theory, Eq.~(\ref{eqn:Jij}), are typically
well consistent with the ones obtained in the more general theory of infinitesimal spin rotations~\cite{JHeisenberg}.
This justifies the perturbative treatment of the transfer integrals and the crystal-field splitting
in the framework of the SE theory.
However, the obtained value of $\alpha_\perp \sim 2 \times 10^{-4}$ ${\rm ps}/{\rm m}$ is very small,
which simply means that the considered spin-current effect is not the main mechanism of the ME coupling in Cr$_2$O$_3$.
This is in line with modern understanding of the ME effect in Cr$_2$O$_3$,
which is known include other important contributions beyond the considered model.
Particularly, the lattice effect is very important~\cite{Bousquet,Malashevich}. Moreover, the orbital magnetization, which is
neglected in the considered half-filled $t_{2g}$ model, can also contribute to $\alpha_\perp$~\cite{Malashevich,Scaramucci}.
We expect that much better agreement with experimental data can be obtained by considering a more general model,
describing the behavior of all Cr $3d$ bands in the basis of Cr $t_{2g}$
\textit{and} $e_g$ Wannier orbitals (see Fig.~\ref{fig.Cr2O3DOS}). For instance, we have found that the DM interactions
are also strongly underestimated in the $t_{2g}$ only model in comparison with the five-orbital model,
where $\boldsymbol{d}_{ij}$ can be computed using Green's function perturbation theory~\cite{SCLR}.
Apparently, the half-filled $t_{2g}$ model is a crude approximation both for DM
exchange interactions and electric polarization
in Cr$_2$O$_3$, and a more relevant microscopic model should include explicitly the effect of the $e_g$ band.
For Cr$_2$O$_3$,
it implies the consideration of several new contributions to the electronic polarization, which are no longer
described by Eq.~(\ref{eqn:Ppartial}) at the half-filling. Below, we will consider several example of $3d^5$ compounds
which are described by a more general model, which explicitly includes both $t_{2g}$ and $e_g$ states at the half filling, and argue
that such model generally provides much better description for the electronic polarization.

\subsection{\label{sec:BiFeO3} Linear Magnetoelectric Effect in BiFeO$_3$}
BiFeO$_3$ is the well known type-I multiferroic, where the FE activity is manly related to the
off-centrosymmetric atomic displacements of Bi and O, while the magnetism originates from the
half-filled $3d$ shell of Fe. The good aspect of BiFeO$_3$ is that the FE and AFM transition temperatures are high
(1100 K and 650 K, respectively), which makes it promising for
practical applications~\cite{MF_review}. In the bulk, due to DM exchange interactions operating in the
noncentrosymmetric $R3c$ structure,
BiFeO$_3$ forms an incommensurate long-periodic spin spiral texture.
The DM interactions overcome the effect magnetocrystalline anisotropy favoring the
conventional G-type antiferromagnetism~\cite{Sosnowska,BiFeO3Jeong,BiFeO3Matsuda}.
Nevertheless, the latter can be stabilized in the
thin films of BiFeO$_3$, where the magnetocrystalline anisotropy can be substantially increased.
An interesting aspect
of the G-phase is that it allows for the linear ME effect, where the electric polarization rises linearly
with the applied magnetic field, whereas in the spin-spiral phase, this effect is averaged to zero by the
spin-spiral modulation. The linear ME coupling $\alpha$ in the BiFeO$_3$ films was first studied experimentally in Ref.~\cite{BiFeO3_Rivera}. However,
the obtained value of $\alpha$
was rather moderate (of the order of $0.51$ ${\rm ps}/{\rm m}$). The interest in this problem has
reemerged again after report
of giant ME coupling, being of the order of 3 ${\rm V}/({\rm cm~Oe})$~\cite{BiFeO3_Wang}.
Even higher value of 24 ${\rm V}/({\rm cm~Oe})$
(corresponding to $3 \times 10^4$ ${\rm ps}/{\rm m}$~\cite{BiFeO3_Stevenson}) was reported later
in the composite films including BiFeO$_3$~\cite{BiFeO3_Lorenz}.

  In this section we will investigate the linear ME effect in BiFeO$_3$.
The computational details and parameters of the model, constructed in the basis of
five Fe$3d$ states near the Fermi level, can be found in the previous publication~\cite{SCLR}.

\begin{figure}[tbp]
\begin{center}
\includegraphics[width=10cm]{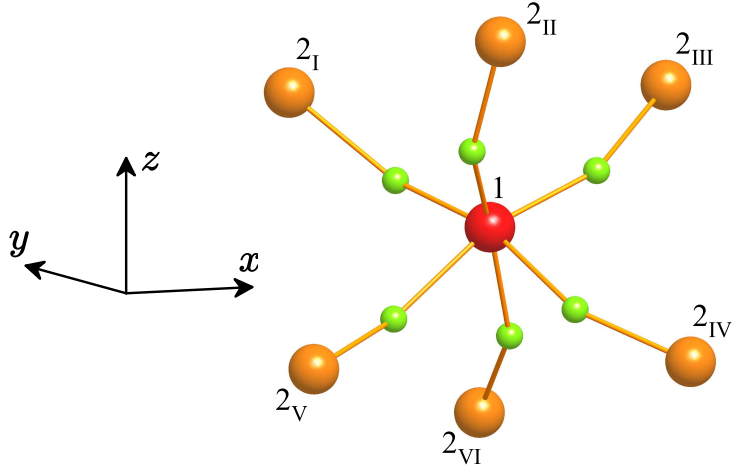}
\end{center}
\caption{(Color online)
Fragment of the crystal structure of BiFeO$_3$: central Fe site of the type 1 is surrounded by
neighboring Fe sites of the type 2 (all are indicated by the big red spheres)
with the notations of their atomic positions.
$\boldsymbol{a}$, $\boldsymbol{b}$, and $\boldsymbol{c}$ denote the directions of trigonal lattice vectors.
The intermediate O atoms are indicated by the small green spheres.}
\label{fig.BiFeO3cluster}
\end{figure}

  The behavior of $\boldsymbol{\cal P}_{ij}$ can be understood on the cluster, where the central Fe site is surrounded by
its six nearest neighbors (Fig.~\ref{fig.BiFeO3cluster}).
In fact, other bonds also produce a sizable contribution to the ME effect in BiFeO$_3$. However, as expected, the
main contribution comes from the nearest neighbors. Moreover, the bonds between Fe sites of the sate type (either 1 or 2)
are ferromagnetically coupled and, therefore, do not contribute to the ME effect (see discussions in Sec.~\ref{sec:Cr2O3}).
For the bond
$\langle 12_{\rm I} \rangle$, corresponding to
$\Delta \boldsymbol{\tau}_{2_{\rm I}1} = (-a,0,\frac{c}{2})$
(where $a = 3.222$ \AA~and $c = 4.625$ \AA~are the rhombohedral lattice parameters),
we obtain
$\boldsymbol{\cal P}_{12_{\rm I}} = (7.88, -3.08, -1.68)$ $\mu {\rm C}/{\rm m}^2$.
The parameters for other bonds can be obtained using the symmetry operations of the group $R3c$,
similar to the DM exchange interactions, which were considered in details in Ref.~\cite{SCLR}.
These parameters $\boldsymbol{\cal P}_{ij}$ are
more than two orders of magnitude larger than the ones obtained in the $t_{2g}$ model for Cr$_2$O$_3$.
Again, due to the threefold rotational symmetry, it holds $\sum_j \boldsymbol{\cal P}_{ij} = 0$.
However, when $\boldsymbol{\cal P}_{ij}$
is combined with $\boldsymbol{\epsilon}_{ji}$ in Eq.~(\ref{eqn:Ppartial2}),
one can expect finite $\boldsymbol{p}_i$.

\begin{figure}[tbp]
\begin{center}
\includegraphics[width=12cm]{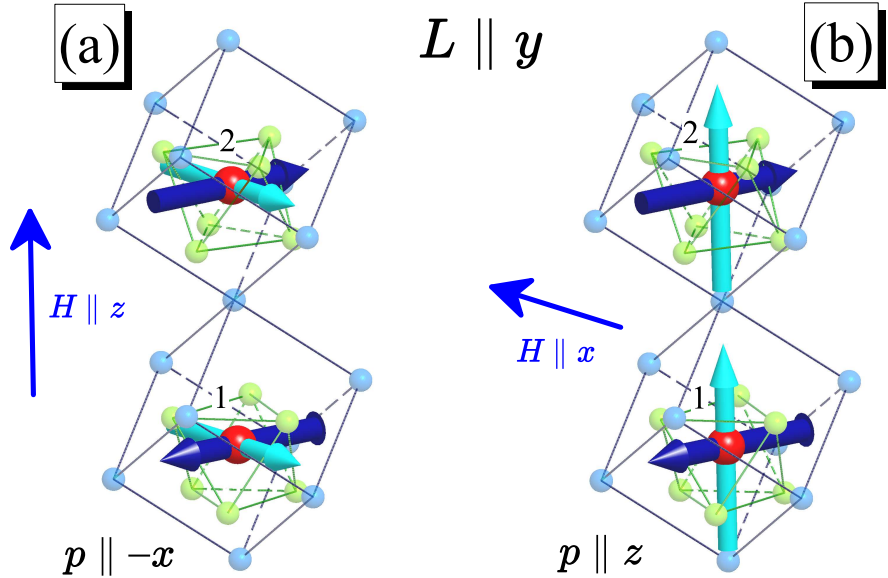}
\end{center}
\caption{(Color online)
Illustration of linear
magnetoelectric effect in BiFeO$_3$: the spin magnetic moments (denoted by dark blue arrows) are parallel to the
$\boldsymbol{y}$ axis.
Then, the external magnetic field along
$\boldsymbol{z}$ or $\boldsymbol{x}$ axes induces the electric polarization (denoted by light cyan arrows)
at both Fe sites of the lattices
along, respectively, $-\boldsymbol{x}$ or $\boldsymbol{z}$.
The Fe atoms are indicated by the big red spheres, the Bi atoms are indicated by
the small blue spheres, and the O atoms are indicated by the small green spheres.}
\label{fig.BiFeO3chain}
\end{figure}

  In our analysis of the ME effect,
we assume that the magnetocrystalline anisotropy confines the spins in the $\boldsymbol{xy}$ plane.
To be specific, we consider here only the case of $\boldsymbol{L} || \boldsymbol{y}$, where
$\boldsymbol{L} = \frac{M}{2} (\boldsymbol{e}_1 - \boldsymbol{e}_2)$ is the AFM order parameter (Fig.~\ref{fig.BiFeO3chain}), but,
due to the $R3c$ symmetry, similar analysis holds also for $\boldsymbol{L} || \boldsymbol{x}$.
Then, we consider the effect of the magnetic field, which cants the spins in the direction of either $\boldsymbol{z}$ or $\boldsymbol{x}$.

  In the first case ($\boldsymbol{H} || \boldsymbol{z}$),
the active component of $\boldsymbol{\cal P}_{ij}$, which is selected by $[ \boldsymbol{e}_1 \times \boldsymbol{e}_2 ]$, is
${\cal P}_{ij}^x$. Then, by combing it with $\boldsymbol{\epsilon}_{ji}$, using the symmetry operation of the $R3c$ group,
and noting that $\frac{\partial}{\partial e^z}[ \boldsymbol{e}_1 \times \boldsymbol{e}_2 ] = 2$,
it is straightforward to show that
$p_1^x \approx - 16 a {\cal P}_{12_{\rm I}}^y /\sqrt{4a^2 + c^2}$, while $p_1^y = p_1^z = 0$.
This NN contribution accounts for 65\% of total $p_1^x = -78.2$ $\mu {\rm C}/{\rm m}^2$,
obtained after summation over all bonds. In the BiFeO$_3$ structure, the Fe sites 1 and 2 are transformed to each other
by the symmetry operation $\{ \hat{m}_y | (0,0,\frac{3c}{2}) \}$, which keeps the sign of $\epsilon^x$, but changes the one of ${\cal P}^x$.
Moreover, in the AFM phase, this transformation flips the directions of spins. Altogether it leads to
the property $p_2^x = p_1^x$ and net electric polarization in the magnetic field.

  In the second case ($\boldsymbol{H} || \boldsymbol{x}$), the active component is ${\cal P}_{ij}^z$, which leads to
the properties:
$p_1^x = p_1^y = 0$ and $p_1^z \approx - 12 c {\cal P}_{12_{\rm I}}^z /\sqrt{4a^2 + c^2}$
(note that in this case $\frac{\partial}{\partial e^x}[ \boldsymbol{e}_1 \times \boldsymbol{e}_2 ] = -2$).
This NN contribution accounts for 60\% of total $p_1^z = 19.7$ $\mu {\rm C}/{\rm m}^2$.
Similar to $\boldsymbol{H} || z$, it is straightforward to show that $p_2^z = p_1^z$, also resulting in finite
ME effect.

  Thus, the induced electronic polarization satisfies the condition
${\bf P} \sim [\boldsymbol{H} \times \boldsymbol{L}]$, being in total agreement with results
of the Ginzburg-Landau theory~\cite{BiFeO3_Popkov}. Finally, we evaluate matrix elements of the ME tensor,
$\alpha_\parallel = \frac{\partial P^z}{\partial H^x}$ and
$\alpha_\perp = \frac{\partial P^x}{\partial H^z}$ (for $\boldsymbol{L} || \boldsymbol{y}$),
using the same procedure as for Cr$_2$O$_3$ and
parameters of
exchange interactions $J_{ij}$
reported in Ref.~\cite{SCLR}, which are consistent with experimental data and
reproduce the experimental value of N\'eel temperature ($T_{\rm N}$).
Then, using the obtained value of $J_0 \approx -241$ meV and $M \approx 5$ $\mu_{\rm B}$,
we will find $ | \alpha_\parallel | = 0.03$ ${\rm ps}/{\rm m}$ and
$| \alpha_\perp | = 0.12$ ${\rm ps}/{\rm m}$. These results are
consistent (at least, by an order of magnitude) with direct calculations of electronic polarization for the model
Hartree-Fock Hamiltonian
without invoking the perturbation theory for the SO coupling, and also the experimental value of
$0.51$ ${\rm ps}/{\rm m}$, reported in Ref.~\cite{BiFeO3_Rivera}.
The giant enhancement of the ME coupling, which was reported in Refs.~\cite{BiFeO3_Wang,BiFeO3_Lorenz},
probably requires additional mechanisms, such as the structural and magnetic reconstruction
in the critical electric field, as was proposed in Refs.~\cite{BiFeO3_Popkov,BiFeO3_Lisenkov,BiFeO3_Kulagin}.

  When the spins lie in the $\boldsymbol{xy}$ plane, there is also an ``intrinsic ME effect'' due to the
FM canting of spins ($\sim 0.5^{\circ}$~\cite{SCLR,EdererSpaldin})
in the direction perpendicular to $\boldsymbol{L}$, which is caused by DM exchange
interactions without any magnetic field.
Below $T_{\rm N}$, it leads to
the polarization change $\Delta P^z$, which
can be estimated, using the obtained values of $\boldsymbol{\cal P}_{ij}$, as $~0.2$ $\mu {\rm C}/{\rm m}^2$.

  Below we will critically examine the main approximations of our theory by considering the DM exchange interactions,
which can be easily computed by using other techniques.
Parameters of DM interactions, obtained in the SE theory for bare $\xi_0 = 53.1$ meV,
$\boldsymbol{d}_{12_{\rm I}} = (0.106,-0.287,0.140)$ meV agree
reasonably well with $\boldsymbol{d}_{12_{\rm I}} = (0.145,-0.418,0.177)$ meV,
derived using more general Green's function perturbation theory method for the same model~\cite{SCLR}.
Both superexchange and Green's function methods
are the first-order theories with respect to the SO coupling. Nevertheless, the Green's function
method does not employ additional approximations, such as the perturbation theory expansion
with respect to the transfer integrals and the crystal-field splitting.
The reasonably good agreement obtained for the DM parameters
demonstrates that such approximations are indeed justifiable.
The conclusion is not so trivial because, for the five-orbital model, the $t_{2g}$-$e_g$ level splitting
in the octahedral environment is not small, being about $1.7$ eV.
Nevertheless, it is smaller than the effective interaction $U \approx 5.8$ eV.
Another important effect is the
polarizability of the electron system by the SO interaction~\cite{SCLR},
which in our case is taken into account only approximately, by
using the effective coupling $\xi = 123$ meV instead of $\xi_0$,
where $\xi$ was derived by fitting results of the SCLR calculations
for matrix elements of the ``screened'' SO interactions with different projections spins.
The ``screened'' SO interaction includes the bare contribution
as well as all the self-consistent change of the Coulomb and exchange
potential, treated on the mean-field level in the first order of the SO coupling.
Thus, the use of $\xi$ instead of $\xi_0$
simply scales the DM parameter $\boldsymbol{d}_{12_{\rm I}}$ by about factor of 2.
Although it captures the main tendency,
it does not describe all details of $\boldsymbol{d}_{12_{\rm I}} = (0.494,-1.450,0.330)$ meV, obtained by combining SCLR
with Green's function perturbation theory, which is the most rigorous method for the evaluation of
DM interactions~\cite{SCLR}.
Thus, our SE theory for the DM interactions and ME coupling is
probably only semi-quantitative one. However, we believe that
it should not change the main conclusions, particularly
regarding the comparison with the experimental data for BiFeO$_3$.

  Finally, we would also like to stress that the phenomenological expression (\ref{eqn:phenomenology})
fails to describe the ME effect in BiFeO$_3$:
for the canted spin structure, inherent to the ME effect, $[\boldsymbol{e}_i \times \boldsymbol{e}_j]$ is the
same for all neighboring bonds surrounding each Fe site. On the other hand, it holds
$\sum_j \boldsymbol{\epsilon}_{ji} = 0$. Thus, no ME effect would be expected if
${\bf P}_{ij} \propto \boldsymbol{\epsilon}_{ji} \times [\boldsymbol{e}_i \times \boldsymbol{e}_j]$,
which is obviously not true.

\subsection{\label{sec:MnWO4} Noncollinear spin order and ferroelectric polarization in MnWO$_4$}
\begin{figure}[h!]
\begin{center}
\includegraphics[width=10cm]{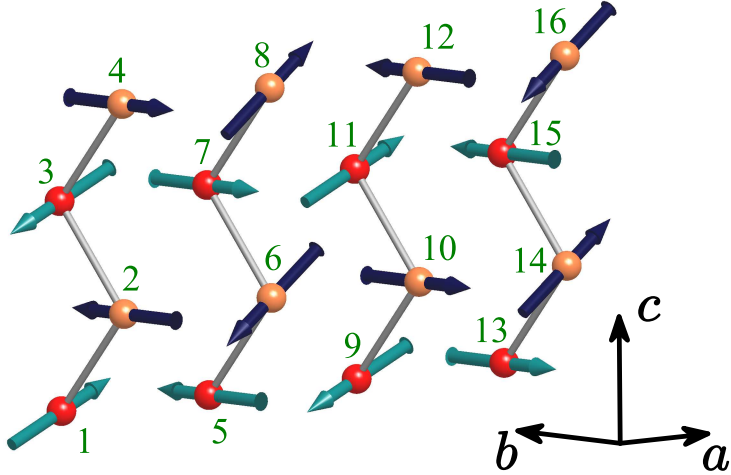}
\end{center}
\caption{\label{fig.spiral}
(Color online) Noncollinear spin-spiral
texture with ${\bf q} = (- \frac{1}{4},\frac{1}{2},\frac{1}{2})$ in MnWO$_4$.
$\boldsymbol{a}$, $\boldsymbol{b}$, and $\boldsymbol{c}$ are the monoclinic translation vectors.
Two Mn sites in the primitive cell of MnWO$_4$, which are transformed to each other by the inversion
operation, are indicated by red (dark) and orange (light) spheres.
}
\end{figure}

MnWO$_4$ has attracted a considerable attention as an example of the spin-spiral magnet, which was
theoretically suggested to be multiferroic~\cite{Heyer},
where this multiferroic behavior
was indeed observed experimentally~\cite{Heyer,Taniguchi,Arkenbout}, and studied in many details after that~\cite{Mitamura,Olabarria1,Olabarria2,Xiao}. Finite polarization was observed in the so-called noncollinear
AF2 phase which is realized in the temperature interval 7.6 K $< T <$ 12.5 K and described by the propagation vector
${\bf q}_{\rm AF2} = (-0.214,\frac{1}{2},0.457)$~\cite{Taniguchi}. The spins rotate in the plane formed by the
monoclinic $\boldsymbol{b}$ axis and one of the axes $\boldsymbol{a}^*$
in the $\boldsymbol{ac}$ plane (see Fig.~\ref{fig.spiral}),
the direction of which
is specified by the single-ion anisotropy.
The electric polarization is parallel
to $\boldsymbol{b}$ axis, but can be realigned along $\boldsymbol{a}$ by applying the external
magnetic field parallel to $\boldsymbol{b}$.
In our previous work (Ref.~\cite{PRB13}) we have
suggested that the FE activity in MnWO$_4$ may be related to the deformation of the spin-spiral texture,
which explicitly breaks the inversion symmetry. The computational details and parameters of the effective low-energy model,
constructed for the half-filled Mn$3d$ bands of MnWO$_4$, can be also found in Ref.~\cite{PRB13}.

  It is interesting to note that, unlike in the magnetoelectric Cr$_2$O$_3$ and BiFeO$_3$,
the direction of polarization in MnWO$_4$
is described by the phenomenological model (\ref{eqn:phenomenology}).
Indeed,
for the spin rotation plane
formed by $\boldsymbol{a}^*$ and $\boldsymbol{b}$, the cross product $[\boldsymbol{e}_i \times \boldsymbol{e}_j]$
is parallel to $\boldsymbol{c}^*$, which is another vector in the $\boldsymbol{ac}$ plane being
perpendicular to $\boldsymbol{a}^*$. Then, for ${\bf q}_{\rm AF2} = (-0.214,\frac{1}{2},0.457)$, there are
two types of neighboring bonds formed by noncollinear spins for which
$\boldsymbol{\epsilon}_{ji} || \boldsymbol{a}$ and $\boldsymbol{\epsilon}_{ji} || \boldsymbol{c}$.
In both cases the expression
${\bf P}_{ij} \propto \boldsymbol{\epsilon}_{ji} \times [\boldsymbol{e}_i \times \boldsymbol{e}_j]$
yields ${\bf P}_{ij} || \boldsymbol{b}$, which agrees with the experimental situation~\cite{Taniguchi}.
Nevertheless, below we will show that such agreement is rather fortuitous and the actual reason behind it
is the specific $P2/c$ symmetry of MnWO$_4$.

\begin{figure}[h!]
\begin{center}
\includegraphics[width=7.1cm]{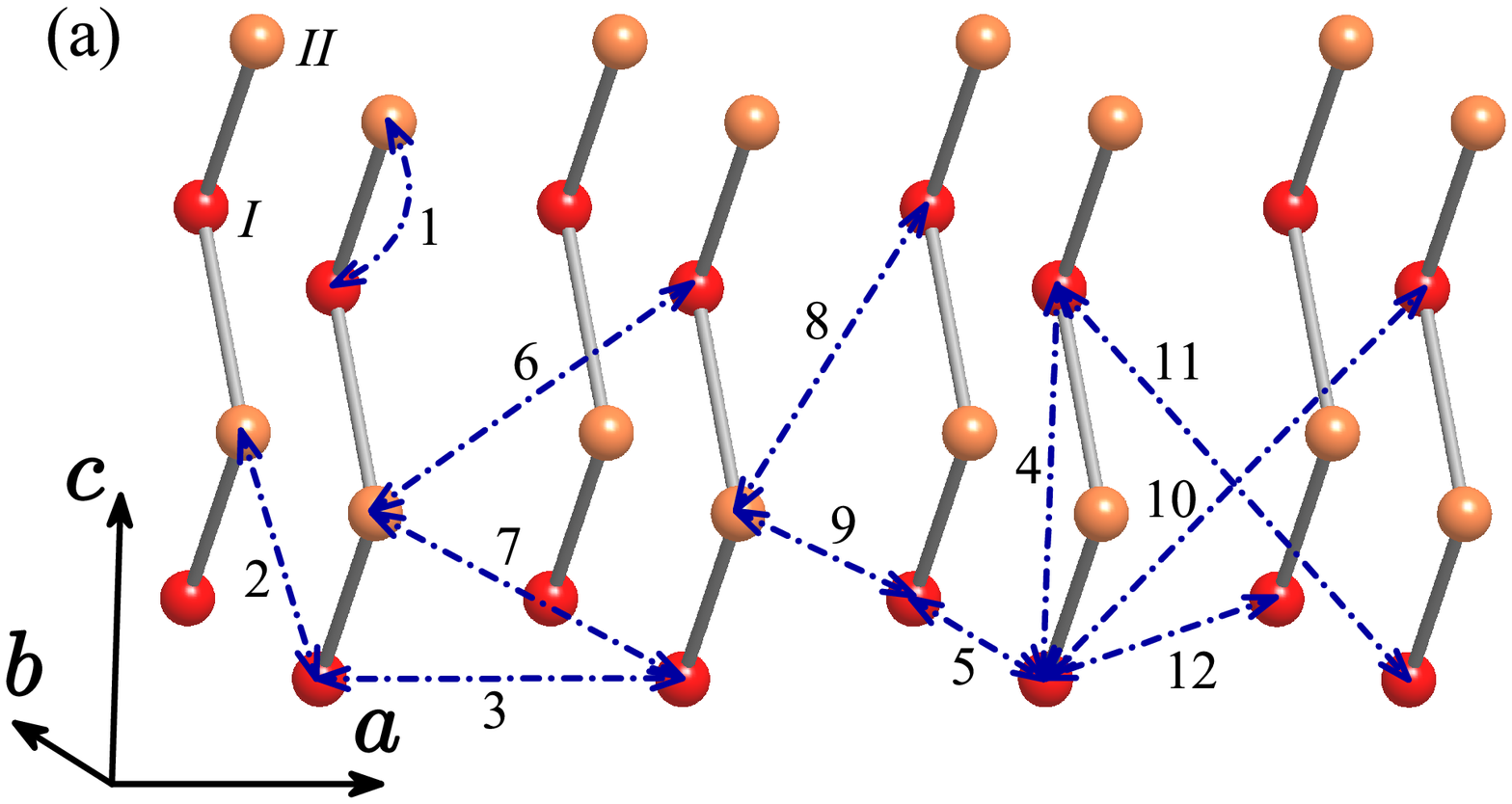}
\includegraphics[width=5.8cm]{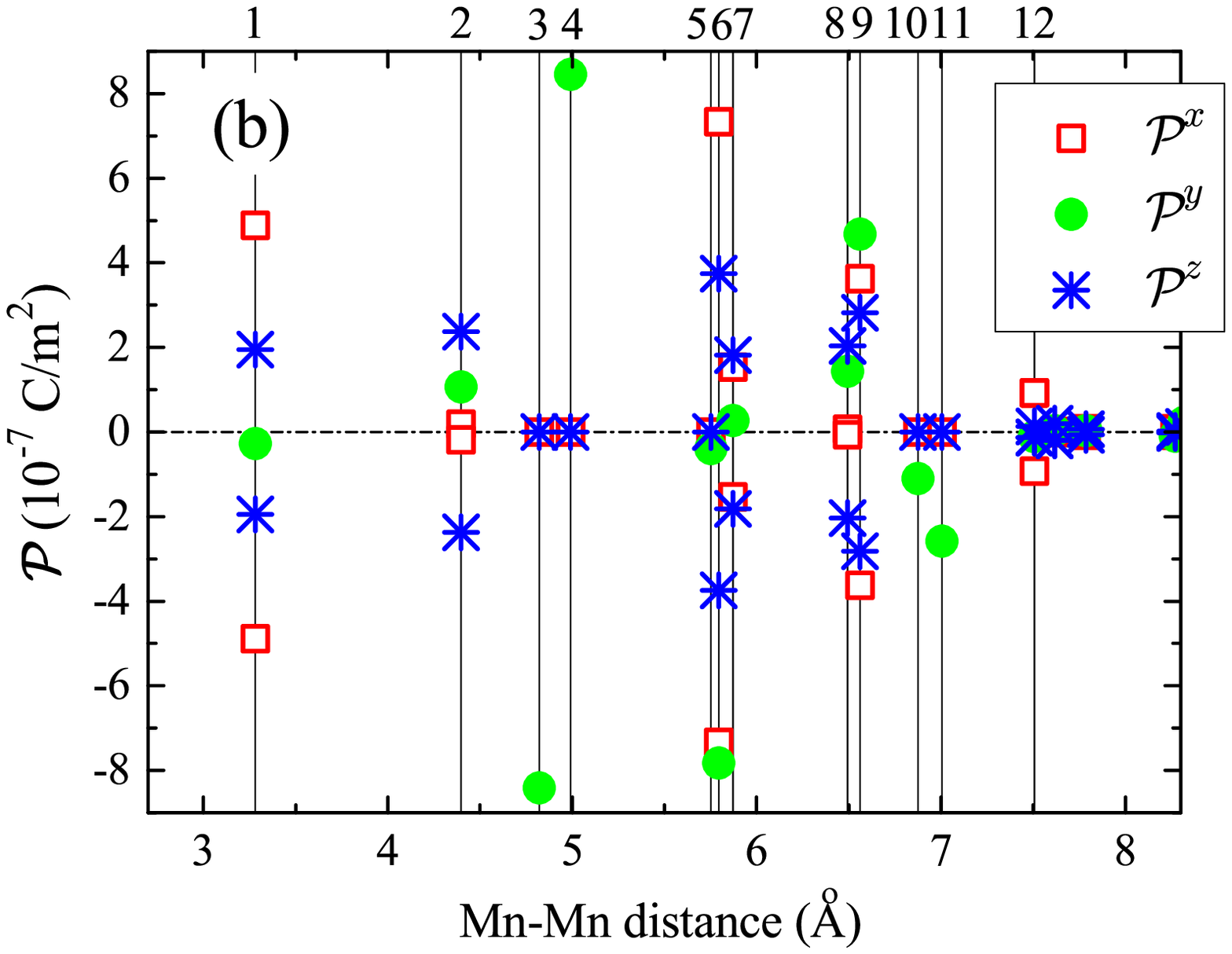}
\end{center}
\caption{\label{fig.MnWO4}
(Color online) (a) Fragment of the crystal structure of MnWO$_4$ with explanation of the bond types
surrounding Mn site $I$ in twelve coordination spheres (other equivalent bonds are not shown).
Two Mn sites in the primitive cell of MnWO$_4$ are denoted as $I$ and $II$.
$\boldsymbol{a}$, $\boldsymbol{b}$, and $\boldsymbol{c}$ are the monoclinic translation vectors.
(b) Distance-dependence of pseudovectors $\boldsymbol{\cal P} = ({\cal P}^x,{\cal P}^y,{\cal P}^z)$
(where $\boldsymbol{y} = \boldsymbol{b}$, $\boldsymbol{z} = \boldsymbol{c}$, and
$\boldsymbol{x}$ is perpendicular to $\boldsymbol{y}$ and $\boldsymbol{z}$), specifying the
electric polarization, in twelve coordination spheres (marked by vertical lines and numbered at the top).
Due to the twofold rotation symmetry about $\boldsymbol{b}$, ${\cal P}^y$ in the equivalent bonds will have
the same signs, while ${\cal P}^x$ and ${\cal P}^z$ have opposite signs, as reflected in the figure.
}
\end{figure}

  The behavior of pseudovectors $\boldsymbol{\cal P}_{ij}$,
reflecting the symmetry properties of MnWO$_4$, is explained in Fig.~\ref{fig.MnWO4}.
The vectors are long-ranged and not restricted by the nearest neighbors. For instance, we have found sizable parameters
spreading up to twelfth coordination sphere.
Similar behavior was found for isotropic exchange interactions
(being in total agreement with the experimental data~\cite{Ye})
and is related to the
long-range character of the transfer integrals~\cite{PRB13}. Due to the twofold rotation about the monoclinic $\boldsymbol{b}$ axis
($\hat{C}_b^2$), which is one of the symmetry operations of the $P2/c$ group (apart from a translation),
${\cal P}^y$ in equivalent bonds will have identical signs, while ${\cal P}^x$ and ${\cal P}^z$ will have opposite signs.
Moreover, if the bond connect two Mn sites of the same type (either $I$ or $II$),
$\hat{C}_b^2$ will transform it to equivalent bond, separated by a translation. Therefore, for this type of bonds we will
have additional condition: ${\cal P}^x = {\cal P}^z =0$.

  Then, we consider the effect of noncollinear spin-spiral texture with
${\bf q} = (- \frac{1}{4},\frac{1}{2},\frac{1}{2})$ (Fig.~\ref{fig.spiral}),
which is close ${\bf q}_{\rm AF2}$ realized in the experimental FE AF2 phase~\cite{Taniguchi}.
We will use this model mainly for numerical estimations, while our symmetry considerations are more general
and valid also for the experimental ${\bf q}_{\rm AF2}$.
First we consider perfect spiral structure texture. The effect of deformation of the spin spiral,
which was proposed in Ref.~\cite{PRB13}, will be investigated below.

  The spin-spiral structure itself is stabilized by competing isotropic exchange interactions~\cite{PRB13}.
However, its spacial orientation depends on the single-ion anisotropy and DM interactions, which also compete
with each other: the former tends to align the spins in the $\boldsymbol{ac}$ plane
(and cant them off the $\boldsymbol{a}$ axis by about $40^\circ$)~\cite{PRB13,Taniguchi},
while the main DM vectors $\boldsymbol{d}_{ij}$ also lie in the $\boldsymbol{ac}$ plane ($\in \boldsymbol{ac}$)~\cite{PRB13}.
Thus, in order to minimize the energy of DM interactions, some of the spins should be parallel to the
$\boldsymbol{b}$ axis ($|| \boldsymbol{b}$), which conflicts with the small single-ion anisotropy.
Moreover, the DM exchange interactions compete with the isotropic ones.
Thus, the situation is indeed very subtle and the magnetic structure is rather fragile.
Nevertheless, this is a very important point because, as we will see in a moment, the spacial orientation of the
spin-spiral plane can control both magnitude \textit{and} direction of the electric polarization.

\begin{figure}[h!]
\begin{center}
\includegraphics[width=10cm]{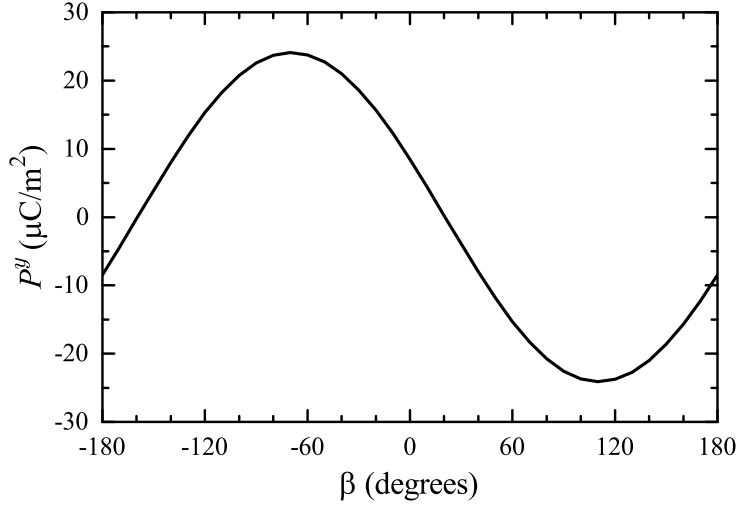}
\end{center}
\caption{\label{fig.Pbeta}
Electric polarization, $P^y$ of the spin-spiral phase of MnWO$_4$ with
${\bf q} = (- \frac{1}{4},\frac{1}{2},\frac{1}{2})$, where half of the spins are parallel to the monoclinic
$\boldsymbol{b}$ axis and another half lies in the $\boldsymbol{ac}$ plane (see Fig.~\ref{fig.spiral}).
$\beta$ is the angle formed by the spins and the monoclinic $\boldsymbol{c}$ axis.
}
\end{figure}

  First, we consider the experimental situation where the spin-spiral plane is formed by the $\boldsymbol{b}$ axis
and one of directions ($\boldsymbol{a}^*$)
in the $\boldsymbol{ac}$ plane~\cite{Taniguchi}.
Then, considering the magnetic structure in Fig.~\ref{fig.spiral}, half of the spins
is parallel to $\boldsymbol{b}$
and another half belongs to $\boldsymbol{ac}$. This means that for the bonds $\langle ij \rangle$ with unparallel spins,
$[\boldsymbol{e}_i \times \boldsymbol{e}_j]$
will also belong to
$\boldsymbol{ac}$ and, therefore, the active components of
$\boldsymbol{\cal P}_{ij}$ are ${\cal P}_{ij}^x$ and ${\cal P}_{ij}^z$. Then, for the
equivalent bond $\langle i'j' \rangle$,
which is obtained from $\langle ij \rangle$ by $\hat{C}_b^2$, we will have the following properties:
${\cal P}_{i'j'}^x =-{\cal P}_{ij}^x$,
${\cal P}_{i'j'}^z =-{\cal P}_{ij}^z$, and
$[\boldsymbol{e}_{i'} \times \boldsymbol{e}_{j'}] = - [\boldsymbol{e}_i \times \boldsymbol{e}_j]$.
The latter property holds because $\hat{C}_b^2$ reverses the direction of propagation of the spin-spiral along
$\boldsymbol{a}$ and $\boldsymbol{c}$. Therefore, if the bond $\langle ij \rangle$ is along the propagation direction,
the bond $\langle i'j' \rangle$ lies in the opposite direction. Thus,
according to Eq.~(\ref{eqn:Ppartial2}), the finite polarization is possible in the direction,
which does not change under $\hat{C}_b^2$, keeping the sign of corresponding projection of the vector
$\boldsymbol{\epsilon}_{ji}$. For the considered geometry of the spin spiral, this direction is
$\boldsymbol{b}$ ($=\boldsymbol{y}$),
in agreement with the experimental data~\cite{Taniguchi}. However,
the absolute value of polarization depends on the orientation of spins in the $\boldsymbol{ac}$ plane.
Indeed, according to Eq.~(\ref{eqn:Ppartial2}),
if $\boldsymbol{e}_i = (0,1,0)$ and $\boldsymbol{e}_j = (\sin \beta,0,\cos \beta)$,
the polarization behaves as
$P_{ij}^y \sim ({\cal P}_{ij}^x \cos \beta - {\cal P}_{ij}^z \sin \beta)$. The dependence of total polarization
$P^y = \sum_j P_{ij}^y$ on $\beta$, obtained using the numerical values of ${\cal P}_{ij}^x$ and ${\cal P}_{ij}^z$,
is displayed in Fig.~\ref{fig.Pbeta}.
Thus, one can conclude the follows: \\
(i) The finite polarization in MnWO$_4$ can be indeed induced by the spiral magnetic order. In this sense,
the conclusion of our previous work~\cite{PRB13} about crucial importance of inhomogeneity
(or deformation) of the spin-spiral order was exaggerated; \\
(ii) The absolute value of $P^y$ strongly depends on the orientation of spins in the $\boldsymbol{ac}$ plane.
The maximal value of about 25 $\mu {\rm C}/{\rm m}^2$ is comparable with experimental
50 $\mu {\rm C}/{\rm m}^2$~\cite{Taniguchi}. However, it does not mean that this maximal value
is realized for the same angle $\beta$, which minimizes the total energy of the system. In fact,
the directions of spins
are controlled by anisotropic interactions, which are small in MnWO$_4$~\cite{PRB13}. Therefore, the situation is very fragile.
This probably explains the large spread of the values of
electric polarizations reported in electronic structure calculations,
which are typically underestimated in comparison with the experimental data~\cite{PRB13,Tian,Shanavas}.

  Then, we consider the situation when all spins lie in the $\boldsymbol{ac}$ plane and also form
the spin spiral with the propagation vector ${\bf q} = (- \frac{1}{4},\frac{1}{2},\frac{1}{2})$.
This behavior was observed experimentally
in the magnetic filed $\boldsymbol{H} || \boldsymbol{b}$, which causes the spin-flop-like transition and
orients the spins in the $\boldsymbol{ac}$ plane, also changing the direction of experimental polarization from
${\bf P} || \boldsymbol{b}$ to mainly ${\bf P} || \boldsymbol{a}$~\cite{Taniguchi}.
In this case,
$[\boldsymbol{e}_i \times \boldsymbol{e}_j]$ is parallel to $\boldsymbol{b}$ and
the active component of $\boldsymbol{\cal P}_{ij}$
is ${\cal P}_{ij}^y$. Then, for two bonds $\langle ij \rangle$ and $\langle i'j' \rangle$,
which are transformed to each other by $\hat{C}_b^2$, we will have: ${\cal P}_{i'j'}^y ={\cal P}_{ij}^y$ and
$[\boldsymbol{e}_{i'} \times \boldsymbol{e}_{j'}] = - [\boldsymbol{e}_i \times \boldsymbol{e}_j]$.
Therefore, the finite polarization is possible in the directions,
which are reversed by $\hat{C}_b^2$. This means that the polarization should lie in the $\boldsymbol{ac}$ plane.
The direction of polarization in the plane is not specified by the symmetry and is the matter of numerical calculations.
Using the numerical values of ${\cal P}_{ij}^y$ (Fig.~\ref{fig.MnWO4}), we obtain
$P^x = 36.6$ $\mu {\rm C}/{\rm m}^2$ and $P^z = 9.4$ $\mu {\rm C}/{\rm m}^2$.
In agreement to the symmetry arguments~\cite{Olabarria2}, our theory also predicts small
polarization along $\boldsymbol{c}$,
which could be verified experimentally. This conclusion is formally consistent with
the phenomenological model (\ref{eqn:phenomenology}).
Nevertheless, we would like to emphasize that the actual reason for such behavior,
both for ${\bf P} || \boldsymbol{b}$ and ${\bf P} \in \boldsymbol{ac}$, is the specific symmetry of MnWO$_4$ and
the existence of the twofold rotation $\hat{C}_b^2$ among symmetry operations of the space group $P2/c$.

  Finally, we discuss the effect of spin-spiral inhomogeneity on the electronic polarization $P^y$ in the ground state,
which was proposed in Ref.~\cite{PRB13}. This inhomogeneity is caused by the competition of
isotropic and DM exchange interactions, which breaks the inversion symmetry and makes two Mn sublattices in MnWO$_4$
inequivalent (shown by different colors in Ref.~\ref{fig.spiral}). Particularly, for the
${\bf q} = (- \frac{1}{4},\frac{1}{2},\frac{1}{2})$ structure, half of the spins will remain parallel to the $\boldsymbol{b}$
axis, while another half will split in two groups, forming different angles $\beta$ with respect to the $\boldsymbol{c}$ axis
($69^\circ$ and $56^\circ$, respectively)~\cite{PRB13}. Then, there will be four types of
Mn sites with distinct neighborhood:
1, 2, 5 and 6 in Fig.~\ref{fig.spiral}, which yield four distinct values of
${\bf P}_i = \sum_j {\bf P}_{ij}$, respectively: $25.3$, $20.0$, $11.8$, and $21.2$ $\mu {\rm C}/{\rm m^2}$.
The total polarization in this case is the average value of these four, yielding $19.6$ $\mu {\rm C}/{\rm m^2}$,
which is consistent with the value of electric polarization $| {\bf P} |$ of homogeneous spin-spiral with the
average $\beta = 61.5^\circ$ (see Fig.~\ref{fig.Pbeta}). Thus, the spin-spiral inhomogeneity does not seem to make a
significant effect on the value of ${\bf P}$ in MnWO$_4$, contrary to manganites, where the polarization is driven
by nonrelativistic double exchange mechanism~\cite{DEpol}.

\subsection{\label{sec:MnO2} Symmetry constraints on the direction of polarization in spin-spiral MnO$_2$}
The rutile ($\beta$-) phase of MnO$_2$ is another interesting example. Due to competing first- and second-neighbor AFM exchange interactions,
it develop the incommensurate spin-spiral order below $T_{\rm N} \approx 92$ K~\cite{MnO2_Yoshimori,MnO2_Sato}.
The spin spiral propagates
along the tetragonal
$\boldsymbol{c}$ axis ($=$$\boldsymbol{z}$). Therefore,
from the viewpoint of spin-current theory,
it could be another potential multiferroic compound~\cite{KNB,Mostovoy}, though has never been considered in this context.
In this section, we will show that the multiferroic effect can be indeed expected in the rutile phase of MnO$_2$.
Moreover, the behavior of
electronic polarization
obeys the phenomenological rule
${\bf P} \propto \boldsymbol{c} \times [\boldsymbol{e}_i \times \boldsymbol{e}_j]$~\cite{KNB,Mostovoy}.
Nevertheless, we will argue that the actual reason behind it is related to the specific $P4_2/mnm$ symmetry of MnO$_2$,
which imposes the symmetry constraints on the properties of $\boldsymbol{\cal P}_{ij}$.

\begin{figure}[tbp]
\begin{center}
\includegraphics[width=10cm]{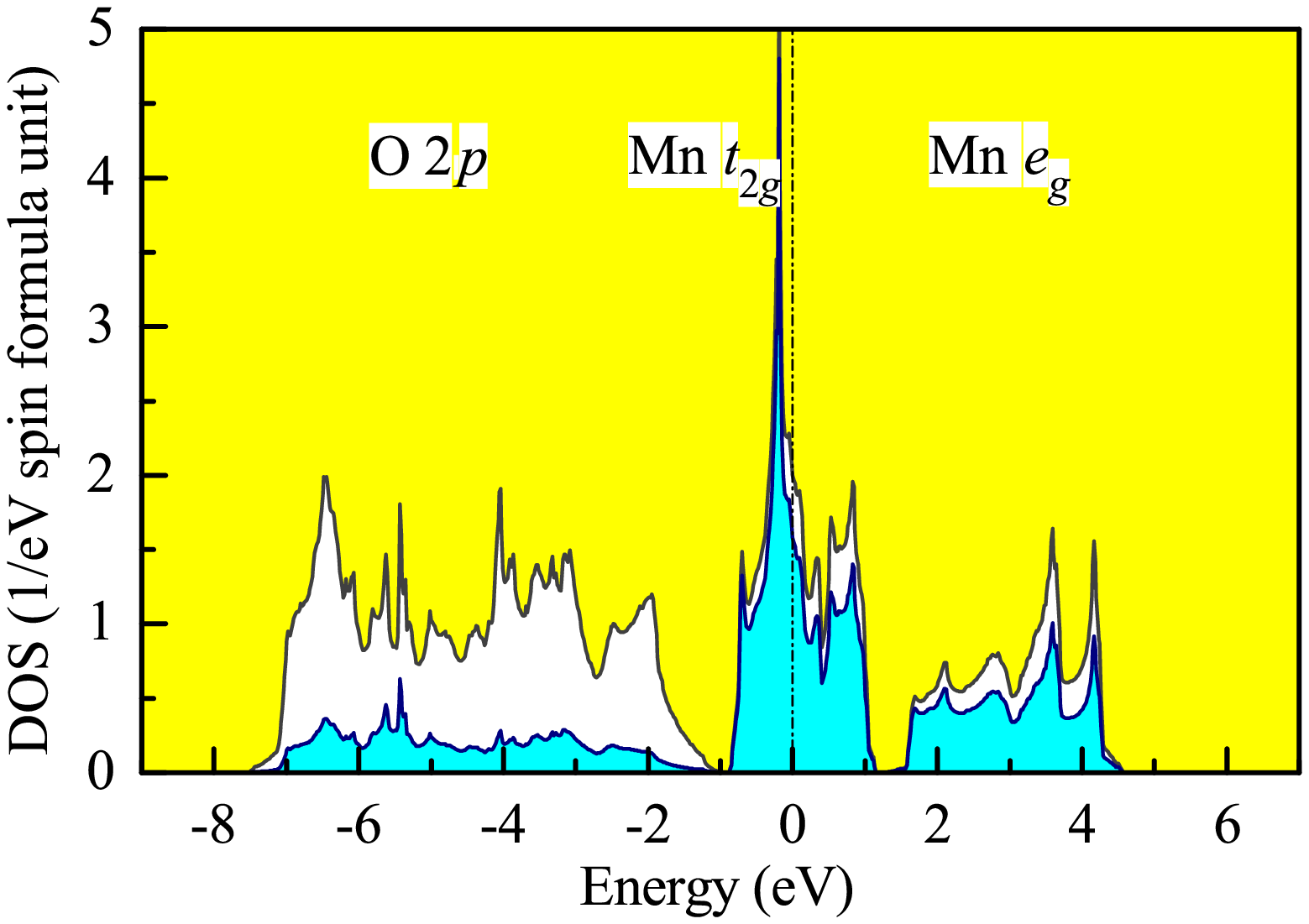}
\end{center}
\caption{(Color online)
Total and partial densities of states of MnO$_2$ in the local density approximation.
The shaded light (blue) area shows contributions of the Mn$3d$ states.
Positions of the main bands are indicated by symbols. The Fermi level is at zero energy (shown by dot-dashed line).}
\label{fig.MnO2DOS}
\end{figure}

  We use the experimental parameters of the crystal structure, reported in Ref.~\cite{MnO2_structure}.
There are two Mn sites in the primitive cell, which are connected by the symmetry operations of the
space group $P4_2/mnm$. Like in Cr$_2$O$_3$, we consider the minimal model comprising of half-filled $t_{2g}$ states
near the Fermi level (Fig.~\ref{fig.MnO2DOS}).
In this case, the crystal-field splitting of $t_{2g}$ levels is pretty large (about $370$ meV).
The Kanamori parameters of screened intraorbital Coulomb and exchange interaction are $3.0$ and $0.72$ eV, respectively.
Other parameters can be found elsewhere~\cite{footnote2}. The isotropic exchange
interactions between first and second neighbors, located at
$(0,0,\pm c )$ and $(\pm \frac{a}{2}, \pm \frac{a}{2}, \pm \frac{c}{2})$
($a$ and $c$ being the tetragonal lattice parameters), are $-16.4$ meV and $-12.3$ meV, respectively.
Like in other considered systems, the theories of SE interactions, Eq.~(\ref{eqn:Jij}), and
infinitesimal spin rotations, Ref.~\cite{JHeisenberg}, give very close values of $J_{ij}$.
The obtained exchange interactions
support the appearance of spin-spiral superstructure with ${\bf q} \approx (0,0,\frac{1}{7})$
(comprising of 7 primitive cells),
in agreement with the
analysis of experimental data~\cite{MnO2_Yoshimori} and results of first-principle calculations~\cite{MnO2_Lim}.
Moreover, the magnetocrystalline anisotropy energy confines the spins in the $\boldsymbol{xy}$ plane.

\begin{figure}[tbp]
\begin{center}
\includegraphics[width=10cm]{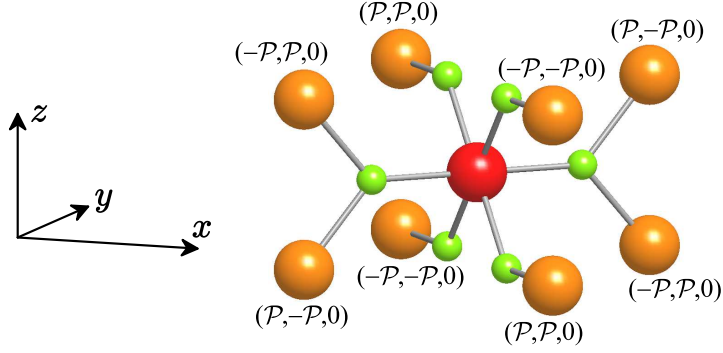}
\end{center}
\caption{(Color online)
Fragment of the crystal structure of MnO$_2$ illustrating the symmetry properties of pseudovectors $\boldsymbol{\cal P}_{ij}$
in eight neighboring bonds, connecting two types of Mn sites.
The Mn atoms are indicated by the big red spheres and the O atoms are indicated by the small green spheres.
The numerical value of parameter ${\cal P}$ is $0.023$ $\mu {\rm C}/{\rm m}^2$.}
\label{fig.MnO2cluster}
\end{figure}

  The $P4_2/mnm$ space group imposes the symmetry constrains on the properties of $\boldsymbol{\cal P}_{ij}$, which
are explained in Fig.~\ref{fig.MnO2cluster}.
The parameters $\boldsymbol{\cal P}_{ij}$ between nearest neighbors vanish identically due to the $mmm$ symmetry in the bonds $(0,0,\pm c)$.
Then, due to the symmetry operations $\{ \hat{C}^4_c| (\frac{a}{2},\frac{a}{2},\frac{c}{2}) \}$ and
$\{ \hat{m}_x| (\frac{a}{2},\frac{a}{2},\frac{c}{2}) \}$ ($\hat{C}^4_c$ being the fourfold rotation about
the tetragonal axis $\boldsymbol{c}$),
transforming the second-neighbor bonds to themselves, the corresponding parameters $\boldsymbol{\cal P}_{ij}$
will have the following form:
$\boldsymbol{\cal P}_{ij} = (\pm {\cal P}, \pm {\cal P},0) $ (see Fig.~\ref{fig.MnO2cluster}). Therefore,
it is straightforward to see that the spin spiral, propagating along $\boldsymbol{c}$ ($= \boldsymbol{z}$) and
rotating in the $\boldsymbol{xy}$ plane, does not induce any polarization because the active component
${\cal P}_{ij}^z$ is identically equals to zero.
For other bonds with lower symmetry, some of ${\cal P}_{ij}^z$ can be finite. However,
the phases of $\boldsymbol{\epsilon}_{ji} {\cal P}_{ij}^z$
alternate for the equivalent types of bonds, again resulting in no net polarization.

  However, when the spins rotate in the $\boldsymbol{yz}$, the active component is ${\cal P}_{ij}^x$, which is finite.
Moreover,
by combining the phases of ${\cal P}_{ij}^x$ with the ones of $\boldsymbol{\epsilon}_{ji}$, it is straightforward to see that
$P^x = P^z = 0$, while $P^y$ can be finite.
Using obtained parameters ${\cal P}_{ij}^x$ we estimate $P^y$ for ${\bf q} \approx (0,0,\frac{1}{7})$
as $2$ $\mu {\rm C}/{\rm cm}^2$.
Similar conclusion holds when the spins rotate in the $\boldsymbol{zx}$ plane.

  Thus, we expect no FE activity in the magnetic ground state of MnO$_2$. However, small polarization can be induced
by rotating the spins to either $\boldsymbol{yz}$ or $\boldsymbol{zx}$ plane. It can be done by applying the external magnetic field
along either $\boldsymbol{x}$ or $\boldsymbol{y}$ axis. Thus, our finding can be verified experimentally.
The result is formally consistent
with the phenomenological expression
${\bf P} \propto \boldsymbol{c} \times [\boldsymbol{e}_i \times \boldsymbol{e}_j]$~\cite{KNB,Mostovoy}.
However, it should be understood that, similar to MnWO$_4$, the
actual reason behind it is the specific symmetry of the rutile phase of MnO$_2$.

\section{\label{sec:conc}Summary and Conclusions}
We have derives an analytical expression for the electronic polarization driven by the SO interaction
in noncollinear magnets. For these purposes we have considered the Berry-phase theory of electric
polarization and applied it to the
Hubbard model at the half filling.
Thus, our analysis is limited by the spin-current mechanism and do not
involve additional complications caused by the orbital degrees of freedom.
Moreover, all derivations are performed in the spirit of the SE theory, which
is valid in the first order of the SO coupling and in the
lowest order of $\hat{t}_{ij}/U$, similar to the
analysis of DM exchange interactions~\cite{Moriya_weakF}.

  We have found that the electric polarization in each bond is
given by Eq.~(\ref{eqn:Ppartial2}), which is substantial revision of the
phenomenological expression (\ref{eqn:phenomenology}). Namely, the electronic polarization
in Eq.~(\ref{eqn:Ppartial2})
explicitly depends on the symmetry of the lattice
(similar to the DM exchange interactions $\boldsymbol{d}_{ij}$~\cite{Dzyaloshinskii_weakF}):
this dependence is described
by the pseudovector $\boldsymbol{\cal P}_{ij}$, which is coupled to the
cross product $[\boldsymbol{e}_i \times \boldsymbol{e}_j]$, depending on the directions of spins.
Thus, this coupling describes
how the symmetry of the lattice interferes with the symmetry of the noncollinear
arrangement of spins. The direction of the polarization itself is specified by the unit
vectors $\boldsymbol{\epsilon}_{ji}$ in the direction connecting two magnetic sites,
which are modulated by the scalar
$\boldsymbol{\cal P}_{ij} \cdot [\boldsymbol{e}_i \times \boldsymbol{e}_j]$.
We argue that,
even though the direction of electric polarization in some spin-spiral magnets can be described by
the phenomenological expression (\ref{eqn:phenomenology}), the actual reason behind it is the
specific symmetry properties of each considered system, which are described by the
pseudovectors $\boldsymbol{\cal P}_{ij}$. Moreover, we have shown that the spin-current mechanism
is much more generic and operates not only in spin-spiral compounds, but also in other types of
noncollinear magnets, where the phenomenological expression (\ref{eqn:phenomenology})
breaks down. Particularly, absolutely the same mechanism may lead to the ME effect
induced by the ferromagnetic canting of spins in the external magnetic field.

  Another important factor, which plays a crucial role
even at the half-filling, is the crystal-field splitting.
We have shown that
without crystal field, both DM exchange interactions and electronic polarization will vanish.
However, the crystal field may have other interesting consequences.
For instance,
it leads to the asphericity in the distribution of the charge density around each
transition-metal site and, if the latter is located not in the centrosymmetric position (like for all considered here compounds),
one can also expect ionic contribution to the electronic polarization,
which can be also derived from the Berry-phase theory,
as was demonstrated recently in Ref.~\cite{Ba2CoGe2O7} for Ba$_2$CoGe$_2$O$_7$.
This is also consistent with the phenomenological analysis by Moriya~\cite{Moriya_pol}, who
expressed the total polarization as the sum of ionic contributions and the ones originating from the bonds.
The ionic contributions
were also evaluated in the present work and found to be at least one order of magnitude smaller than the
``anomalous'' pair contributions, which are given by Eq.~(\ref{eqn:Ppartial2}) and related to fine
details of the electronic structure~\cite{Picozzi}.

  Using parameters of the effective Hubbard model, derived from the first-principles electronic structure calculations,
we have evaluated the spin-current contribution to the electronic polarization for the series of
ME materials (Cr$_2$O$_3$ and BiFeO$_3$)
and multiferroics compounds with the spin-spiral structure (MnWO$_4$ and MnO$_2$).
We have shown that, although Eq.~(\ref{eqn:Ppartial2}) excellently reproduces the symmetry properties of
polarization, its numerical values can be severely underestimated, depending on the material. Particularly, the
largest disagreement was found for the ME effect in Cr$_2$O$_3$, which suggest the importance of
other (lattice and orbital) contributions, in agreement with the previous finding~\cite{Bousquet,Malashevich,Scaramucci}.

  We have also clarified the microscopic origin of FE activity in the spin-spiral phase of MnWO$_4$: although the
spin spiral in this case is deformed by competing isotropic and antisymmetric DM exchange interactions,
which explicitly breaks the inversion symmetry~\cite{PRB13}, this deformation seems to have a minor effect
on the value of electronic polarization. The main contribution to the polarization comes from the spin-current term,
given by Eq.~(\ref{eqn:Ppartial2}), which also describes the change of the direction of polarization,
depending on the spacial orientation of the spin spiral.

  Finally, we have predicted the FE activity in the rutile phase of MnO$_2$ when the spin spiral is rotated
our of the tetragonal $\boldsymbol{xy}$ plane.


\begin{thebibliography}{99}
\bibitem{footnote2}
All model parameters are available upon request.
\end{thebibliography}
\end{document}